\documentclass[fleqn,a4paper]{article}
\usepackage{a4wide}
\usepackage{amsmath}
\usepackage{amssymb}
\usepackage{comment}
\usepackage{hyphenat}
\usepackage{multicol}
\usepackage{graphicx}
\usepackage{amsfonts}
\usepackage{bm}
\usepackage{graphicx}
\usepackage{xcolor}
\usepackage{natbib}
\usepackage[linesnumbered,ruled]{algorithm2e}

\newcommand{\qed}{\hfill$\Box$}

\title{A robust approach to model-based classification based on trimming and constraints \\ \normalsize Semi-supervised learning in presence of outliers and label noise}
\author{Andrea Cappozzo\footnote{Department of Statistics and Quantitative Methods, University of Milano-Bicocca, \texttt{a.cappozzo@campus.unimib.it, francesca.greselin@unimib.it}} \and  Francesca Greselin\footnotemark[\value{footnote}] \and Thomas Brendan Murphy\footnote{School of Mathematics \& Statistics and Insight Research Centre, University College Dublin, \texttt{brendan.murphy@ucd.ie}}}

\begin{document}
\maketitle

\abstract{
In a standard classification framework a set of trustworthy learning data are employed to build a decision rule, with the final aim of classifying unlabelled units belonging to the test set. Therefore, unreliable labelled observations, namely outliers and data with incorrect labels, can strongly undermine the classifier performance, especially if the training size is small. The present work introduces a robust modification to the Model-Based Classification framework, employing impartial trimming and constraints on the ratio between the maximum and the minimum eigenvalue of the group scatter matrices. The proposed method effectively handles noise presence in both response and exploratory variables, providing reliable classification even when dealing with contaminated datasets.
A robust information criterion is proposed for model selection. Experiments on real and simulated data, artificially adulterated, are provided to underline the benefits of the proposed method. 
}

\section{Introduction}\label{intro}
In statistical learning, we define classification as the task of assigning group memberships to a set of unlabelled observations. Whenever a labelled sample (i.e., the training set) is available, the information contained in such dataset is exploited to classify the remaining unlabelled observations (i.e., the test set), either in a supervised or in a semi-supervised manner, depending whether the information contained in the test set are included in building the classifier \citep[e.g.][]{PaulDMcNicholas16}. Either way, the presence of unreliable data points can be detrimental for the classification process, especially if the training size is small \citep{Zhu}.

Broadly speaking, noise is anything that obscures the relationship between the attributes and the class membership  \citep{Hickey1996}. In a classification context, \cite{Wu:1995:KAD:527260} distinguishes between two types of noise: attribute noise and class noise. The former is related to contamination in the exploratory variables, that is when observations present unusual values on their predictors; whereas the latter refers to samples whose associated labels are wrong. \cite{Zhu} and the recent work of \cite{Prati2018} offer an extensive review on the topic and the methods that have been proposed in the literature to deal with attribute noise and class noise, respectively. Generally, three main approaches can be employed when building a classifier from a noisy dataset: cleaning the data, modeling the noise and using robust estimators of model parameters \citep{Bouveyron2009a}.

The approach presented in this paper is based on a robust estimation of a Gaussian mixture model with parsimonious structure, to account for both attribute and label noise. Our conjecture is that the contaminated observations would be the least plausible units under the robustly estimated model: the corrupted subsample will be revealed by detecting those observations with the lowest contributions to the associated likelihood. Impartial trimming \citep{Gordaliza1991, Gordaliza1991a, Cuesta_Albertos1997} is employed for robustifying the parameter estimates, being a well established technique to treat mild and gross outliers in the clustering literature \citep{Garc??a-Escudero2010} and here used, for the first time, to additionally account for label noise in a classification framework. A semi-supervised approach is developed, where information contained in both labelled and unlabelled samples is combined for improving the classifier performance and for defining a data-driven method to identify outlying observations possibly present in the test set.

The rest of the manuscript is organized as follows. A brief review on model-based discriminant analysis and classification is given in Section \ref{sec:DA},  Section \ref{sec:REDDA} introduces the robust updating classification rules, covering the model formulation, inference aspects and model selection. Simulation studies to compare the method introduced in Section \ref{sec:REDDA} with other popular classification methods are reported in Section  \ref{sec:sim_study}. Finally, in Section \ref{sec:Honey} our proposal is employed in performing classification and adulteration detection in a food authenticity context, dealing with contaminated samples of Irish honey. Concluding notes and further research directions are outlined in Section  \ref{sec:conclusion}. The proof of Proposition 1, details on the parameters values for simulation study II and efficient algorithms for enforcing the eigen-ratio constraint for different patterned models are deferred respectively to appendices A, B and C.
%That is, employing Impartial Trimming observations whose likelihood is lowest given the current model will not be accounted for in the parameter estimation \citep{Gordaliza1991, Gordaliza1991a, Cuesta_Albertos1997}.
% both in terms of uncertain labels and contamination in the explanatory variables, 
%We propose to use impartial trimming for robustifying the estimates, since it is a well estabilshed tool to treat mild and gross outliers and it is here novelly use to treat label noise             
\section{Model-Based Discriminant Analysis and Classification} \label{sec:DA}
In this Section we review the main concepts of supervised classification based on mixture models, with particular focus on Eigenvalue Decomposition Discriminant Analysis and its semi-supervised formulation, as introduced in \cite{Dean2006}. This approach is the basis of the novel robust semi-supervised classifier introduced in Section \ref{sec:REDDA}.
\subsection{Eigenvalue Decomposition Discriminant Analysis} \label{EDDA}
%The aim of supervised classification is to build a decision rule 
Model-based discriminant analysis \citep{mclachlan2004discriminant, Fraley2002} is a probabilistic approach for supervised classification, in which a classifier is built from a complete set of learning observations $\{(\mathbf{x}_1, \mathbf{l}_1),\ldots, (\mathbf{x}_N, \mathbf{l}_N)\}$; where $\mathbf{x}_n$ and $\mathbf{l}_{n}$, $n=1,\ldots,N$, are independent realizations of random vectors $\mathcal{X}\in \mathbb{R}^p$ and $\mathcal{G}\in \{1,\ldots,G\}$, respectively. That is, $\mathbf{x}_n$ denotes a $p$-variate observation and $\mathbf{l}_{n}$ its associated class label, such that $l_{ng}=1$ if observation $n$ belongs to group $g$ and $0$ otherwise, $g=1,\ldots, G$. Considering a Gaussian framework, the probabilistic mechanism that is assumed to have generated the data is as follows:%where $\mathbf{x}_n$, $n=1,\ldots,N$ are independent realizations of a continuous random vector $\mathcal{X}\in \mathbb{R}^p$ and $\mathbf{l}_{n}$ are the associated class labels, such that $l_{ng}=1$ if observation $n$ belongs to group $g$ and $0$ otherwise, $g=1,\ldots, G$. Considering a Gaussian framework, the probabilistic mechanism that is assumed to have generated the data is as follows: %for genuine observations is represented as follows:
%and $\mathbf{Y}$ are $N \times p$ and $M \times p$ matrices whose rows are realizations of a continuous random vector $\mathcal{V}\in \mathbb{R}^p$, $\mathbf{\overline{l}}$ is a $N \times G$ class label matrix such that $\overline{l}_{ng}=1$ if observation $n$ belongs to group $g$ and $0$ otherwise.
%Model-based discriminant analysis is a probabilistic approach for supervised classification of continuous data in which the data generating process for genuine observations is represented as follows:
\begin{align} \label{dgp_EDDA}
\begin{split}
%\mathcal{G} \sim \prod_{g=1}^G \tau_g^{\mathbb{I}\{\mathcal{G}=g\}}\\
\mathcal{G} \sim Mult_G (1; \tau_1,\ldots, \tau_G)\\
\mathcal{X}|\mathcal{G}=g \sim \mathcal{N}_p(\boldsymbol{\mu}_g, \boldsymbol{\Sigma}_g)
\end{split}
\end{align}
where $\mathcal{G}$ is multinomially distributed with $\tau_g$ probability of observing class $g$ and the conditional density of $\mathcal{X}$ given $\mathcal{G}$ is multivariate normal with mean vector $\boldsymbol{\mu}_g$ and variance covariance matrix $\boldsymbol{\Sigma}_g$. Therefore, the joint density of $(\mathbf{x}_n, \mathbf{l}_n)$ is given by:
\begin{equation}
f(\mathbf{x}_n,\mathbf{l}_n; \boldsymbol{\Theta}) = \prod_{g=1}^G \left[ \tau_g \phi(\mathbf{x}_n; \boldsymbol{\mu}_g, \boldsymbol{\Sigma}_g) \right]^{l_{ng}}
\end{equation}
where $\phi(\cdot; \boldsymbol{\mu}_g, \boldsymbol{\Sigma}_g)$ denotes the multivariate normal density and $\boldsymbol{\Theta}$ represents the collection of parameters to be estimated, $\boldsymbol{\Theta}= \{ \tau_1, \ldots,  \tau_G, \boldsymbol{\mu}_1, \ldots, \boldsymbol{\mu}_G, \boldsymbol{\Sigma}_1, \ldots, \boldsymbol{\Sigma}_G \}$. Discriminant  analysis makes use of data with known labels to  estimate  model  parameters  for creating a classification rule. The trained classifier is subsequently employed for assigning a set of unlabelled observations $\mathbf{y}_m$, $m=1,\ldots, M$ to the class $g$ with the associated highest posterior probability:
\begin{equation}\label{MAP}
z_{mg}=\mathbb{P}(\mathcal{G}=g|\mathcal{X}=\textbf{y}_m)=\frac{\tau_g \phi(\mathbf{y}_m; \boldsymbol{\mu}_g, \boldsymbol{\Sigma}_g)}{\sum_{j=1}^G\tau_j \phi(\mathbf{y}_m; \boldsymbol{\mu}_j, \boldsymbol{\Sigma}_j)}.
\end{equation}
using the maximum a posteriori (MAP) rule. The afore-described framework is widely employed in classification tasks, thanks to its probabilistic formulation and well-established efficacy.

The number of parameters in the component variance covariance matrices grows quadratically with the dimension $p$. Thus, \cite{Bensmail1996} introduced a parsimonious parametrization proposing to enforce additional assumptions on the matrices structure, based on the eigen-decomposition of \cite{Banfield1993} and \cite{Celeux1995}:
\begin{equation} \label{sigma_dec}
\boldsymbol{\Sigma}_g=\lambda_g\boldsymbol{D}_g\boldsymbol{A}_g\boldsymbol{D}^{'}_g
\end{equation}
where $\boldsymbol{D}_g$ is an orthogonal matrix of eigenvectors, $\boldsymbol{A}_g$ is a diagonal matrix such that $|\boldsymbol{A}_g|=1$ and $\lambda_g=|\boldsymbol{\Sigma}_g|^{1/p}$. This elements correspond respectively to the orientation, shape and volume (alternatively called scale) of the different Gaussian components. Allowing each parameter in \eqref{sigma_dec} to be equal or different across groups, \cite{Bensmail1996} define a family of 14 patterned models, listed in Table \ref{EDDA_model}. Such class of models is particularly flexible, as it includes very popular classification methods like Linear Discriminant Analysis and Quadratic Discriminant Analysis as special cases for the EEE and VVV models, respectively \citep{Hastie1996}. %Such family of patterned models is known as Eigenvalue Decomposition Discriminant Analysis (EDDA) and it is implemented in the \texttt{mclust R} package \citep{Fop2016}.
Eigenvalue Decomposition Discriminant Analysis (EDDA) is implemented in the \texttt{mclust R} package \citep{Fop2016}.    

\begin{table}
\centering
\caption{Nomenclature, covariance structure and number of free parameters in $\boldsymbol{\Sigma}_1,\ldots,\boldsymbol{\Sigma}_G$: $\gamma$ denotes the number of parameters related to the orthogonal rotation and $\delta$ the number of parameters related to the eigenvalues. The last column indicates whether the eigenvalue-ratio (ER) constraint is required.}\label{EDDA_model}       % Give a unique label
% For LaTeX tables use
\begin{tabular}{lcccc}
\hline\noalign{\smallskip}
 Model & $\boldsymbol{\Sigma}_g$&$\gamma$& $\delta$ & ER \\
\noalign{\smallskip}\hline\noalign{\smallskip}
 EII & $\lambda \boldsymbol{I}$ & - & $1$ & Not required \\ 
  VII & $\lambda_g \boldsymbol{I}$ & - & $G$ & Required\\ 
 EEI & $\lambda \boldsymbol{A}$ & - & $p$ & Not required\\ 
   VEI & $\lambda_g \boldsymbol{A}$ & - & $G + p -1$ & Required\\
    EVI & $\lambda \boldsymbol{A}_g$ & - & $Gp-(G-1)$  & Required\\
     VVI & $\lambda_g \boldsymbol{A}_g$ & - & $Gp$  & Required\\
EEE & $\lambda\boldsymbol{D}\boldsymbol{A}\boldsymbol{D}^{'}$ & $p (p-1)/2$ & $p$  & Not required\\
  VEE & $\lambda_g \boldsymbol{D}\boldsymbol{A}\boldsymbol{D}^{'}$ & $p (p-1)/2$ & $G + p -1$  & Required\\
EVE & $\lambda\boldsymbol{D}\boldsymbol{A}_g\boldsymbol{D}^{'}$ & $p (p-1)/2$ & $Gp-(G-1)$  & Required \\
EEV & $\lambda\boldsymbol{D}_g\boldsymbol{A}\boldsymbol{D}^{'}_g$ & $G p (p-1)/2$ & $p$  & Not required\\
VVE & $\lambda_g\boldsymbol{D}\boldsymbol{A}_g\boldsymbol{D}^{'}$ & $p (p-1)/2$ & $Gp$   & Required\\
 VEV & $\lambda_g\boldsymbol{D}_g\boldsymbol{A}\boldsymbol{D}^{'}_g$ & $G p (p-1)/2$ & $G + p -1$ & Required\\
 EVV & $\lambda\boldsymbol{D}_g\boldsymbol{A}_g\boldsymbol{D}^{'}_g$ & $G p (p-1)/2$ & $Gp-(G-1)$ & Required\\
 VVV & $\lambda_g\boldsymbol{D}_g\boldsymbol{A}_g\boldsymbol{D}^{'}_g$ & $G p (p-1)/2$ & $Gp$   & Required\\
\noalign{\smallskip}\hline
\end{tabular}
\end{table}
 
\subsection{Updating Classification Rules} \label{upclass}
Exploiting the assumption that the data generating process outlined in \eqref{dgp_EDDA} is the same for both labelled and unlabelled observations, \cite{Dean2006} propose to include also the data whose memberships are unknown in the parameter estimation. That is, information about group structure that may be contained in both labelled and unlabelled samples is combined in order to improve the classifier performance, in a semi-supervised manner. %Previous work related to updating model-based classification includes \citep{McLachlan1975, Oneill1978} and more recently \citep{Frame2007, Mcnicholas2009, Murphy2012} and 	\citep{Toher2011}. For an overview of semi-supervised methods, please refer to \citep{Chapelle2006}.

Under the  framework defined in Section \ref{EDDA}, and given the set of available information $\{(\mathbf{x}_n, \mathbf{l}_n)|n=1,\ldots,N \}\cup \{\mathbf{y}_m|m=1,\ldots, M\}$, the \textit{observed log-likelihood} is
\begin{align}\label{obs_ll}
\begin{split}
\ell(\boldsymbol{\tau}, \boldsymbol{\mu}, \boldsymbol{\Sigma}| \mathbf{X}, \mathbf{Y},\mathbf{l})&=
\sum_{n=1}^N \sum_{g=1}^Gl_{ng} \log{\left[\tau_g \phi(\mathbf{x}_n; \boldsymbol{\mu}_g, \boldsymbol{\Sigma}_g)\right]} +\\
&+ \sum_{m=1}^M \log{\left[\sum_{g=1}^G\tau_g \phi(\mathbf{y}_m; \boldsymbol{\mu}_g, \boldsymbol{\Sigma}_g)\right]}
\end{split}
\end{align}
in which both labelled and unlabelled samples are accounted for in the likelihood definition. Treating the (unknown) labels $z_{mg}$, $m=1,\ldots, M$, $g=1,\ldots, G$ as missing data and including them in the likelihood specification defines the so called \textit{complete-data log-likelihood:}
\begin{align}\label{com_ll}
\begin{split}
\ell_C(\boldsymbol{\tau}, \boldsymbol{\mu}, \boldsymbol{\Sigma}| \mathbf{X}, \mathbf{Y},\mathbf{l}, \mathbf{z})&=
\sum_{n=1}^N \sum_{g=1}^G l_{ng} \log{\left[\tau_g \phi(\mathbf{x}_n; \boldsymbol{\mu}_g, \boldsymbol{\Sigma}_g)\right]} +\\
&+ \sum_{m=1}^M \sum_{g=1}^G z_{mg} \log{\left[\tau_g \phi(\mathbf{y}_m; \boldsymbol{\mu}_g, \boldsymbol{\Sigma}_g)\right]}
\end{split}
\end{align} 
Maximum likelihood estimates for \eqref{obs_ll} are obtained through the EM algorithm \citep{Dempster1977}, iteratively computing the expected value for the unknown labels given the current set of parameter estimates (E-Step), and employing \eqref{com_ll} to find maximum likelihood estimates for the unknown parameters (M-Step). %and to compute the expected value for the unknown labels given the current set of parameter estimates (E-Step).
The unlabelled data are then classified according to $\hat{z}_{mg}$, using the MAP. The updating classification rules was demonstrated to give improved classification performance over the classical model-based discriminant analysis in some food authenticity applications, particularly when the training size is small. An implementation of this can be found in the \texttt{upclass R} package \citep{russell2014upclass}.

\section{Robust Updating Classification Rules}\label{sec:REDDA}
We introduce here a Robust modification to the Updating Classification Rule described in Section \ref{upclass}, with the final aim of developing a classifier whose performance is not affected by contaminated data, either in the form of label noise and outlying observations.
\subsection{Model Formulation}
The main idea of the proposed approach is to employ techniques originated in the branch of robust statistics to obtain a model-based classifier in which parameters are robustly estimated and outlying observations identified. We are interested in providing a method that jointly accounts for noise on response and exploratory variables, where the former might be present in the labelled set and the latter in both the labelled and unlabelled sets. %Our conjecture is that it is reasonable to expect that adulterated observations would be the least plausible under the robust estimated model: the illegal subsample will be revealed by detecting those observations with the lowest contributions to the associated likelihood.
We propose to modify the log-likelihood in \eqref{obs_ll} with a \textit{trimmed mixture log-likelihood} \citep{Neykov2007} and to employ impartial trimming and constraints on the covariance matrices for achieving both robust parameter estimation and identification of the unreliable sub-sample. 
%That is, in employing Impartial Trimming observations whose likelihood is lowest given the current model will not be accounted for in the parameter estimation \citep{Gordaliza1991, Gordaliza1991a, Cuesta_Albertos1997}.
%Given the distinct structure of the likelihoods associated to the labelled and unlabelled sets, impartial trimming is induced in two different manners with respect to the considered set, accounting for the possible label noise that might be present in the labelled sample (see Section \ref{EM_alg} for details). Following the same notation introduced in Section \ref{EDDA}, we aim at maximizing the \textit{trimmed observed data log-likelihood}:
Impartial trimming is enforced by considering the distinct structure of the likelihoods associated to the labelled and unlabelled sets, accounting for the possible label noise that might be present in the labelled sample (see Section \ref{EM_alg} for details). Following the same notation introduced in Section \ref{EDDA}, we aim at maximizing the \textit{trimmed observed data log-likelihood}:
\begin{align} \label{trim_ll}
\begin{split}
\ell_{trim}(\boldsymbol{\tau}, \boldsymbol{\mu}, \boldsymbol{\Sigma}| \mathbf{X}, \mathbf{Y},\mathbf{\text{l}})&=
\sum_{n=1}^N \zeta(\mathbf{x}_n)\sum_{g=1}^G\text{l}_{ng} \log{\left[\tau_g \phi(\mathbf{x}_n; \boldsymbol{\mu}_g, \boldsymbol{\Sigma}_g)\right]} + \\
&+ \sum_{m=1}^M \varphi(\mathbf{y}_m)\log{\left[\sum_{g=1}^G\tau_g \phi(\mathbf{y}_m; \boldsymbol{\mu}_g, \boldsymbol{\Sigma}_g)\right]}
\end{split}
\end{align}
where \(\zeta(\cdot)\), \(\varphi(\cdot)\) are 0-1 trimming indicator functions, that express whether observation \(\mathbf{x}_n\) and \(\mathbf{y}_m\) are trimmed off or not. A fixed fraction \(\alpha_{l}\) and \(\alpha_{u}\) of observations, belonging to the labelled and unlabelled set respectively, is unassigned by setting \(\sum_{n=1}^N \zeta(\mathbf{x}_n)=\lceil N(1-\alpha_{l})\rceil\) and \(\sum_{m=1}^M \varphi(\mathbf{y}_m)=\lceil M(1-\alpha_{u})\rceil\). In this way, the less plausible samples under the currently estimated model are tentatively trimmed out at each step of the iterations that leads to the final estimate. The \textit{labelled trimming level} \(\alpha_{l}\) and the \textit{unlabelled trimming level} \(\alpha_{u}\) account for possible adulteration in both sets.  At the end of the iterations, a value of $\zeta(\mathbf{x}_n)=0$ or $\varphi(\mathbf{y}_m)=0$ corresponds to identify $\mathbf{x}_n$ or $\mathbf{y}_m$, respectively, as unreliable observations. Notice that impartial trimming automatically deals with both class noise and attribute noise, as observations that suffer from either noise structure will give low contribution to the associated likelihood.

Maximization of \eqref{trim_ll} is carried out via the EM algorithm, in which an appropriate Concentration Step \citep{Driessen1999} is performed in both labelled and unlabelled sets at each iteration to enforce the impartial trimming. In addition, we protect the parameter estimation from spurious solutions, that may arise whenever one component of the mixture fits a random pattern in the data. We consider the eigenvalue-ratio restriction:%the labelled sample size for a given group is particularly small, considering an eigenvalues-ratio restriction:
\begin{equation} \label{eigen_contr}
\textbf{M}_n/\textbf{m}_n\leq c
\end{equation}
where 
$\textbf{M}_n=\max_{g=1\ldots G}\max_{l=1\ldots p} d_{lg}$ and 
$\textbf{m}_n=\min_{g=1\ldots G}\min_{l=1\ldots p} d_{lg}$,
with \(d_{lg}\), $l=1,\ldots, p$ being the eigenvalues of the matrix
\(\boldsymbol{\Sigma}_g\) and $c \geq1$ being a fixed constant	 \citep{Ingrassia2004}. Constraint \eqref{eigen_contr} simultaneously controls differences between groups and departures from sphericity, by forcing the relative length of the axes of the equidensity ellipsoids, based on the multivariate normal distribution, to be smaller than $\sqrt{c}$ \citep{Garcia-Escudero2014}. Notice that the constraint in \eqref{eigen_contr} is still needed whenever either the shape or the volume is free to vary across components \citep{Garcia-Escudero}, that is for all models in Table \ref{EDDA_model} that present ``Required'' entry in the ER column. %  for which the Volume and/or Shape columns have ``Variable'' entries.
The considered approach is the (semi)-supervised version of the methodology proposed in \cite{Dotto2019}, which is framed in a completely unsupervised scenario. Feasible and computationally efficient algorithms for enforcing the eigen-ratio constraint for different patterned models are reported in the Appendix C.  

\subsection{Estimation Procedure} \label{EM_alg}
The EM algorithm for obtaining Maximum Trimmed Likelihood Estimates of the robust updating classification rule involves the following steps:

\begin{itemize}
\item
\emph{Robust Initialization:} set \(k=0\). Employing only the labelled data, we obtain robust starting values for the mean vector $\boldsymbol{\mu}_g$ and covariance matrix $\boldsymbol{\Sigma}_g$ of the multivariate normal density for each group $g$, $g=1,\ldots,G$, employing the following procedure:
  \begin{enumerate}
\item For each class $g$, draw a random $(p + 1)$-subset $J_g$ and compute its empirical mean $\hat{\boldsymbol{\mu}}^{(0)}_g
$ and variance covariance matrix $\hat{\boldsymbol{\Sigma}}^{(0)}_g$ according to the considered parsimonious structure. %(mclust::mstep)
This procedure yields better initial subsets than drawing random $\lceil N(1-\alpha_{l})\rceil$-subsets directly, because the probability of drawing an outlier-free $(p + 1)$-subset is much higher than that of drawing an outlier-free $\lceil N(1-\alpha_{l}) \rceil$-subset \citep{Hubert2018}.
\item Set
\begin{align*}
\begin{split}
\hat{\boldsymbol{\theta}}&=\{ \hat{\tau}_1,\ldots,\hat{\tau}_G, \hat{\boldsymbol{\mu}}_1,\ldots,\hat{\boldsymbol{\mu}}_G,
\hat{\boldsymbol{\Sigma}}_1,\ldots,\hat{\boldsymbol{\Sigma}}_G\}=\\
&=\{\hat{\tau}_1^{(0)},\ldots,\hat{\tau}_G^{(0)}, \hat{\boldsymbol{\mu}}_1^{(0)},\ldots,\hat{\boldsymbol{\mu}}_G^{(0)},
\hat{\boldsymbol{\Sigma}}^{(0)}_1,\ldots,\hat{\boldsymbol{\Sigma}}_G^{(0)} \}
\end{split}
\end{align*}
where $\hat{\tau}_1^{(0)}=\ldots=\hat{\tau}_G^{(0)}=1/G$.
%\(\hat{\boldsymbol{\theta}}=\{ \hat{\tau}_1,\ldots,\hat{\tau}_G, \hat{\boldsymbol{\mu}}_1,\ldots,\hat{\boldsymbol{\mu}}_G,
%\hat{\boldsymbol{\Sigma}}_1,\ldots,\hat{\boldsymbol{\Sigma}}_G\}=\{\hat{\tau}_1^{(0)},\ldots,\hat{\tau}_G^{(0)}, \hat{\boldsymbol{\mu}}_1^{(0)},\ldots,\hat{\boldsymbol{\mu}}_G^{(0)},
%\hat{\boldsymbol{\Sigma}}^{(0)}_1,\ldots,\hat{\boldsymbol{\Sigma}}_G^{(0)} \} \) where $\hat{\tau}_1^{(0)}=\ldots=\hat{\tau}_G^{(0)}=1/G$. %(no needed in practice, just to be coherent with the notation).
\item %\emph{Concentration Step:}
For each $\mathbf{x}_n$, $n=1,\ldots,N$, compute the conditional density
\begin{equation}\label{cond_dens}
f(\mathbf{x}_n|l_{ng}=1; \hat{\boldsymbol{\theta}})=\phi\left(\mathbf{x}_n; \hat{\boldsymbol{\mu}}_g, \hat{\boldsymbol{\Sigma}}_g \right) \:\:\:\:\: g=1,\ldots, G.
\end{equation}
\(\lfloor N\alpha_{l} \rfloor \%\) of the samples with lowest value of \eqref{cond_dens} are temporarily discarded as possible outliers, namely label noise and/or attribute noise. That is, $\zeta(\mathbf{x}_n)=0$ for such observations.
\item The parameter estimates are updated,
  based on the non-discarded observations:
  \begin{align}
  \hat{\tau}_g=\frac{\sum_{n=1}^N \zeta(\mathbf{x}_n)l_{ng}}{\lceil N(1-\alpha_{l})\rceil}\:\:\:\:\: g=1,\ldots, G\\
  \hat{\boldsymbol{\mu}}_g=\frac{\sum_{n=1}^N \zeta(\mathbf{x}_n)l_{ng}\mathbf{x}_n}{\sum_{n=1}^N\zeta(\mathbf{x}_n)l_{ng}}\:\:\:\:\: g=1,\ldots, G.
    \end{align}
  Estimation of $\boldsymbol{\Sigma}_g$ depends on the considered patterned model, details are given in \cite{Bensmail1996}.
  \item Iterate $3-4$ until the $\lfloor N\alpha_{l} \rfloor$ discarded observations are exactly the same on two consecutive iterations, then stop (usually, $\leq3$ iterations are required).
\end{enumerate}
The procedure described in steps $1-5$ is performed $\texttt{nsamp}$ times, and the parameter estimates $\hat{\boldsymbol{\theta}}^R$ that lead to the highest value of the objective function %$\hat{\boldsymbol{\theta}}^R= \argmax_{i=1,\ldots,\texttt{nsamp}} \ell_{trim}(\hat{\boldsymbol{\tau}}, \hat{\boldsymbol{\mu}}, \hat{\boldsymbol{\Sigma}}| \mathbf{X}, \mathbf{\text{l}})= \sum_{n=1}^N \zeta(\mathbf{x}_n)\sum_{g=1}^G\text{l}_{ng} \log{\left[\hat{\tau}_g \phi(\mathbf{x}_n; \hat{\boldsymbol{\mu}}_g, \hat{\boldsymbol{\Sigma}}_g)\right]}$ are retained.
$\ell_{trim}(\hat{\boldsymbol{\tau}}, \hat{\boldsymbol{\mu}}, \hat{\boldsymbol{\Sigma}}| \mathbf{X}, \mathbf{\text{l}})= \sum_{n=1}^N \zeta(\mathbf{x}_n)\sum_{g=1}^G\text{l}_{ng} \log{\left[\hat{\tau}_g \phi(\mathbf{x}_n; \hat{\boldsymbol{\mu}}_g, \hat{\boldsymbol{\Sigma}}_g)\right]}$, out of $\texttt{nsamp}$ repetitions, are retained. The afore-described procedure stems from the ideas of the FastMCD algorithm of \cite{Driessen1999}, % for obtaining a computationally efficient robust estimator of multivariate location and scatter,
here adapted for dealing with parsimonious structures in the covariance matrices. Retaining $\hat{\boldsymbol{\theta}}^R$ as final estimates leads to a fully supervised robust model-based method, called REDDA hereafter (see Section \ref{sim_setup}).   
 % The starting values are obtained via standard Eigenvalue Decomposition Discriminant Analysis (see Section \ref{EDDA}). That is, find \(\hat{\tau}_g^{(0)}\), \(\hat{\boldsymbol{\mu}}_g^{(0)}\) and
%\(\hat{\boldsymbol{\Sigma}}_g^{(0)}\) $g=1,\ldots,G$ using only the labelled data. This can be performed using \texttt{MclustDA} routine in the \texttt{mclust} package.
Then, if the selected patterned model allows for heteroscedastic $\boldsymbol{\Sigma}_g$ and \eqref{eigen_contr} is not satisfied, constrained maximization is enforced, see Appendix C for details.
\item
   \emph{EM Iterations:} denote by \(\hat{\boldsymbol{\theta}}^{(k)}=\{ \hat{\tau}_1^{(k)},\ldots,\hat{\tau}_G^{(k)}, \hat{\boldsymbol{\mu}}_1^{(k)},\ldots,\hat{\boldsymbol{\mu}}_G^{(k)},
\hat{\boldsymbol{\Sigma}}^{(k)}_1,\ldots,\hat{\boldsymbol{\Sigma}}_G^{(k)} \} \)  the parameter estimates at the $k$-th iteration of the algorithm.

\begin{itemize}
\item
  \emph{Step 1 - Concentration}: the trimming procedure is implemented by
  discarding the \(\lfloor N\alpha_{l} \rfloor\) observations \(\mathbf{x}_n\) with smaller values of
  \begin{equation} \label{trim_l}
  D\left(\mathbf{x}_n; \hat{\boldsymbol{\theta}}^{(k)} \right)=\prod_{g=1}^G \left[ \phi \left(\mathbf{x}_n; \hat{\boldsymbol{\mu}}^{(k)}_g, \hat{\boldsymbol{\Sigma}}^{(k)}_g \right)\right]^{l_{ng}} \:\:\:\:\: n=1,\ldots,N
    \end{equation}
  and discarding the \(\lfloor M\alpha_{u}\rfloor\) observations \(\mathbf{y}_m\)  with smaller values of
  \begin{equation} \label{trim_u}
D\left(\mathbf{y}_m; \hat{\boldsymbol{\theta}}^{(k)}\right)=\sum_{g=1}^G \hat{\tau}^{(k)}_g \phi \left(\mathbf{y}_m; \hat{\boldsymbol{\mu}}^{(k)}_g, \hat{\boldsymbol{\Sigma}}^{(k)}_g \right) \:\:\:\:\: m=1,\ldots,M.
 \end{equation}  

\item
  \emph{Step 2 - Expectation}: for each non-trimmed observation \(\mathbf{y}_m\)
  compute the posterior probabilities
  \begin{equation}
\hat{z}_{mg}^{(k+1)}=\frac{\hat{\tau}^{(k)}_g \phi \left(\mathbf{y}_m; \hat{\boldsymbol{\mu}}^{(k)}_g, \hat{\boldsymbol{\Sigma}}^{(k)}_g \right)}{D\left(\mathbf{y}_m; \hat{\boldsymbol{\theta}}^{(k)}\right)} \:\:\:\:\: g=1,\ldots, G; \:\:\:\: m=1,\ldots, M.
\end{equation}
\item
  \emph{Step 3 - Constrained Maximization}: the parameter estimates are updated,
  based on the non-discarded observations and the current estimates for the unknown labels:
  \begin{align}
  \hat{\tau}_g^{(k+1)}=\frac{\sum_{n=1}^N \zeta(\mathbf{x}_n)l_{ng}+ \sum_{m=1}^M \varphi(\mathbf{y}_m)\hat{z}_{mg}^{(k+1)}}{\lceil N(1-\alpha_{l})\rceil+\lceil M(1-\alpha_{u})\rceil}\:\:\:\:\: g=1,\ldots, G\\
  \hat{\boldsymbol{\mu}}_g^{(k+1)}=\frac{\sum_{n=1}^N \zeta(\mathbf{x}_n)l_{ng}\mathbf{x}_n+\sum_{m=1}^M\varphi(\mathbf{y}_m)\hat{z}_{mg}^{(k+1)}\mathbf{y}_m}{\sum_{n=1}^N\zeta(\mathbf{x}_n)l_{ng}+\sum_{m=1}^M\varphi(\mathbf{y}_m)\hat{z}_{mg}^{(k+1)}}\:\:\:\:\: g=1,\ldots, G.
    \end{align}
  Estimation of $\boldsymbol{\Sigma}_g$ depends on the considered patterned model and on the eigenvalues-ratio constraint. Details are given in \cite{Bensmail1996} and, if \eqref{eigen_contr} is not satisfied, in Appendix C.
\item

  \emph{Step 4 - Convergence of the EM algorithm}: check for algorithm convergence (see Section \ref{convergence}). If convergence has not been reached, set \(k=k+1\) and repeat steps 1-4.
 \end{itemize}  
\end{itemize}

%Notice that how the trimming step is enforced changes between the labelled and unlabelled observations. Particularly, in the labelled set samples whose conditional density $f(\mathbf{x}_n|l_{ng}=1; \hat{\boldsymbol{\theta}}^{(k)})=\phi\left(\mathbf{x}_n; \hat{\boldsymbol{\mu}}^{(k)}_g, \hat{\boldsymbol{\Sigma}}^{(k)}_g \right)$ is lowest are tentatively trimmed out from the estimation procedure. The reason for trimming the conditional density being that if the group probabilities were to be included (i.e., performing trimming on the joint density) groups with small prior probabilities would likely be trimmed off completely for large enough values of $\alpha_l$. Note that with \eqref{trim_l} we are both discriminating label noise (i.e., observations that are likely to belong to the mixture model but whose associated label is wrong) and outliers. On the other hand, having no prior information on the group membership of the unlabelled units, in \eqref{trim_u} we are temporarily discarding those observations $\mathbf{y}_m$ with lowest marginal density $f(\mathbf{y}_m; \hat{\boldsymbol{\theta}}^{(k)})=\sum_{g=1}^G \hat{\tau}^{(k)}_g \phi \left(\mathbf{y}_m; \hat{\boldsymbol{\mu}}^{(k)}_g, \hat{\boldsymbol{\Sigma}}^{(k)}_g \right)$
Notice how the trimming step differs between the labelled and unlabelled observations. We implicitly assume that a label in the training set conveys a sound meaning about the presence of a class of objects. Therefore, in the labelled set, we opted for trimming the samples with lowest conditional density $f(\mathbf{x}_n|l_{ng}=1; \hat{\boldsymbol{\theta}}^{(k)})=\phi\left(\mathbf{x}_n; \hat{\boldsymbol{\mu}}^{(k)}_g, \hat{\boldsymbol{\Sigma}}^{(k)}_g \right)$. The alternative choice of considering the joint density $f(\mathbf{x}_n, l_{ng}; \hat{\boldsymbol{\theta}}^{(k)})=\prod_{g=1}^G \left[ \tau_g \phi(\mathbf{x}_n; \boldsymbol{\mu}_g, \boldsymbol{\Sigma}_g) \right]^{l_{ng}}$ is instead prone to trim off completely groups with small prior probability $\tau_g$ for large enough value of $\alpha_l$, and should be discarded. Note that with \eqref{trim_l} we are both discriminating label noise (i.e., observations that are likely to belong to the mixture model but whose associated label is wrong) and outliers. In the unlabelled set, on the other hand, trimming is based on the marginal density $f(\mathbf{y}_m; \hat{\boldsymbol{\theta}}^{(k)})=\sum_{g=1}^G \hat{\tau}^{(k)}_g \phi \left(\mathbf{y}_m; \hat{\boldsymbol{\mu}}^{(k)}_g, \hat{\boldsymbol{\Sigma}}^{(k)}_g \right)$, having no prior information on the group membership of the samples.

Once convergence is reached, the estimated values \(\hat{z}_{mg}\) provide a classification for the unlabelled observations \(\mathbf{y}_m\), assigning observation $m$ into group $g$ if $\hat{z}_{mg}>\hat{z}_{mg^{'}}$ for all $g^{'}\neq g$. Final values of \(\zeta(\mathbf{x}_n)=0\), and \(\varphi(\mathbf{y}_m)=0\),   classify $\mathbf{x}_n$ and $\mathbf{y}_m$ respectively, as outlying observations.

The routines for estimating the robust updating classification rules have been written in \texttt{R} language \citep{RCoreTeam2018}: the source code is available at https://github.com/AndreaCappozzo/rupclass.
%the authors upon request, and an \texttt{R} package is currently under development. %is twofold: we want both to consider the 
The estimation procedure detailed in this Section implies the monotonicity of the algorithm, according to:\\ \\
\textbf{Proposition 1:} If the values \(\zeta(\mathbf{x}_n)\), \(\varphi(\mathbf{y}_m)\), $n=1,\ldots,N$, $m=1,\ldots,M$ are kept fixed, the EM algorithm described in Section \ref{EM_alg} implies $\ell_{trim}(\hat{\boldsymbol{\theta}}^{(k+1)}| \mathbf{X}, \mathbf{Y},\mathbf{\text{l}}) \geq \ell_{trim}(\hat{\boldsymbol{\theta}}^{(k)}| \mathbf{X}, \mathbf{Y},\mathbf{\text{l}})$ at any $k$.\\ \\
%The proof is reported in Appendix A. Furthermore, the existence of a maximum of \eqref{trim_ll} and its consistency may be derived, under general conditions, adapting the theoretical results presented in \cite{Genovese2014} and \cite{Garcia-Escudero2015}.
The proof is reported in Appendix A. Furthermore, our estimation procedure reduces possible incorrect modes of the optimization function (spurious maximizers) and offers a constructive way to obtain a maximizer $\hat{\theta}_n$ for the sample problem, that converges to the global maximizer for the population, see \cite{Genovese2014} and \cite{Garcia-Escudero2015}.
%$\ell_{trim}(\boldsymbol{\tau}^{(k+1)}, \boldsymbol{\mu}^{(k+1)}, \boldsymbol{\Sigma}^{(k+1)}| \mathbf{X}, \mathbf{Y},\mathbf{\text{l}}) \geq \ell_{trim}(\boldsymbol{\tau}^{(k)}, \boldsymbol{\mu}^{(k)}, \boldsymbol{\Sigma}^{(k)}| \mathbf{X}, \mathbf{Y},\mathbf{\text{l}})$ at any $k$.\\

\subsection{Convergence Criterion} \label{convergence}
We assess whether the EM algorithm has reached convergence evaluating at each iteration how close the trimmed log-likelihood is to its estimated asymptotic value, using the Aitken acceleration \citep{Aitken1926}:
\begin{equation}
a^{(k)}=\frac{\ell_{trim}^{(k+1)}-\ell_{trim}^{(k)}}{\ell_{trim}^{(k)}-\ell_{trim}^{(k-1)}}
\end{equation}
where $\ell_{trim}^{(k)}$ is the trimmed observed data log-likelihood from iteration $k$. The asymptotic estimate of the trimmed log-likelihood at iteration $k$ is given by \citep{Bohning94}:
\begin{equation}
\ell_{\infty_{trim}}^{(k)} = \ell_{trim}^{(k)} + \frac{1}{1-a^{(k)}}\left(\ell_{trim}^{(k+1)}-\ell_{trim}^{(k)}\right).
\end{equation}
The EM algorithm is considered to have converged when $|\ell_{\infty_{trim}}^{(k)}-\ell_{trim}^{(k)}|<\varepsilon$; a value of $\varepsilon=10^{-5}$ has been chosen for the experiments reported in the next Sessions.
  \begin{figure} 
% Use the relevant command to insert your figure file.
% For example, with the graphicx package use
\centering
 \includegraphics[width=\textwidth, height=7cm]{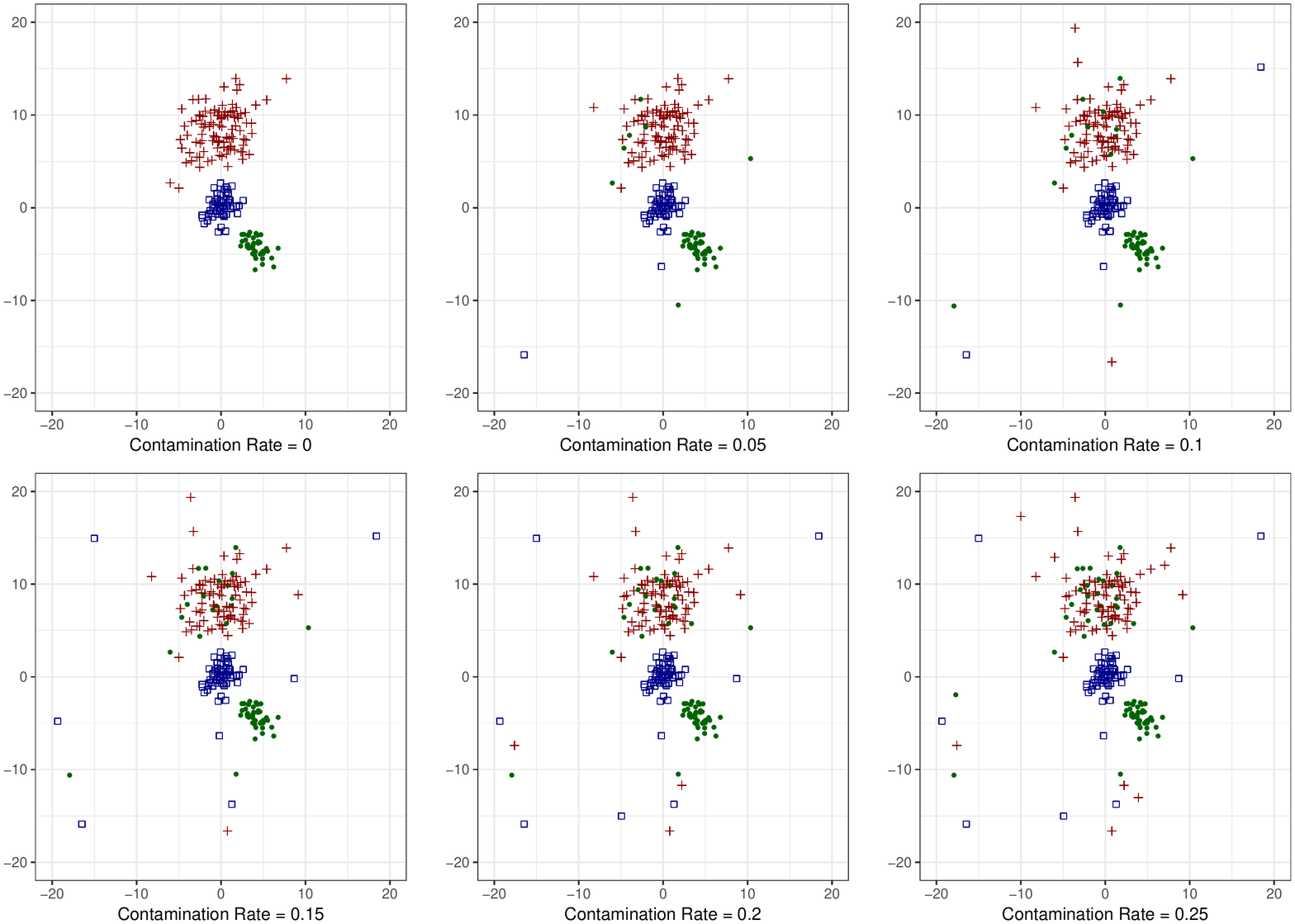}
% figure caption is below the figure
\caption{Simulated data considering the Simulation Setup described in Section \ref{sim_setup}, varying Contamination Rate $\eta$}
\label{fig:cont}     % Give a unique label
\end{figure} 
\subsection{Model Selection} \label{sec:RBIC}
A robust likelihood-based criterion is employed for choosing the best model among the 14 patterned covariance structures listed in Table \ref{EDDA_model} and a reasonable value for the constraint $c$ in \eqref{eigen_contr}:
\begin{equation} \label{RBIC}
RBIC = 2 \ell_{trim}(\hat{\boldsymbol{\tau}}, \hat{\boldsymbol{\mu}}, \hat{\boldsymbol{\Sigma}}) - v_{XXX}^c \log\left(\lceil N(1-\alpha_{l})\rceil + \lceil M(1-\alpha_{u})\rceil\right)
\end{equation}
where $\ell_{trim}(\hat{\boldsymbol{\tau}}, \hat{\boldsymbol{\mu}}, \hat{\boldsymbol{\Sigma}})$ denotes the maximized trimmed observed data log-likelihood and $v_{XXX}^c$ a penalty term whose definition is:
\begin{equation}
v_{XXX}^c = Gp + G-1 + \gamma + (\delta-1)\left(1-\frac{1}{c}\right)+1.
\end{equation}
That is, $v_{XXX}^c$ depends on the total number of parameters to be estimated: $\gamma$ and $\delta$ for every $XXX$ patterned model are given in Table \ref{EDDA_model}. It also accounts for the trimming levels and for the eigen-ratio constraint $c$, according to \cite{Cerioli2018}. Note that, when $c\rightarrow +\infty$ and $\alpha_l=\alpha_u=0$, \eqref{RBIC} is the Bayesian Information Criterion \citep{Schwarz1978}. 
%\subsection{Subsection title}
\begin{figure}
 \centering 
% Use the relevant command to insert your figure file.
% For example, with the graphicx package use
 \includegraphics[scale=.5]{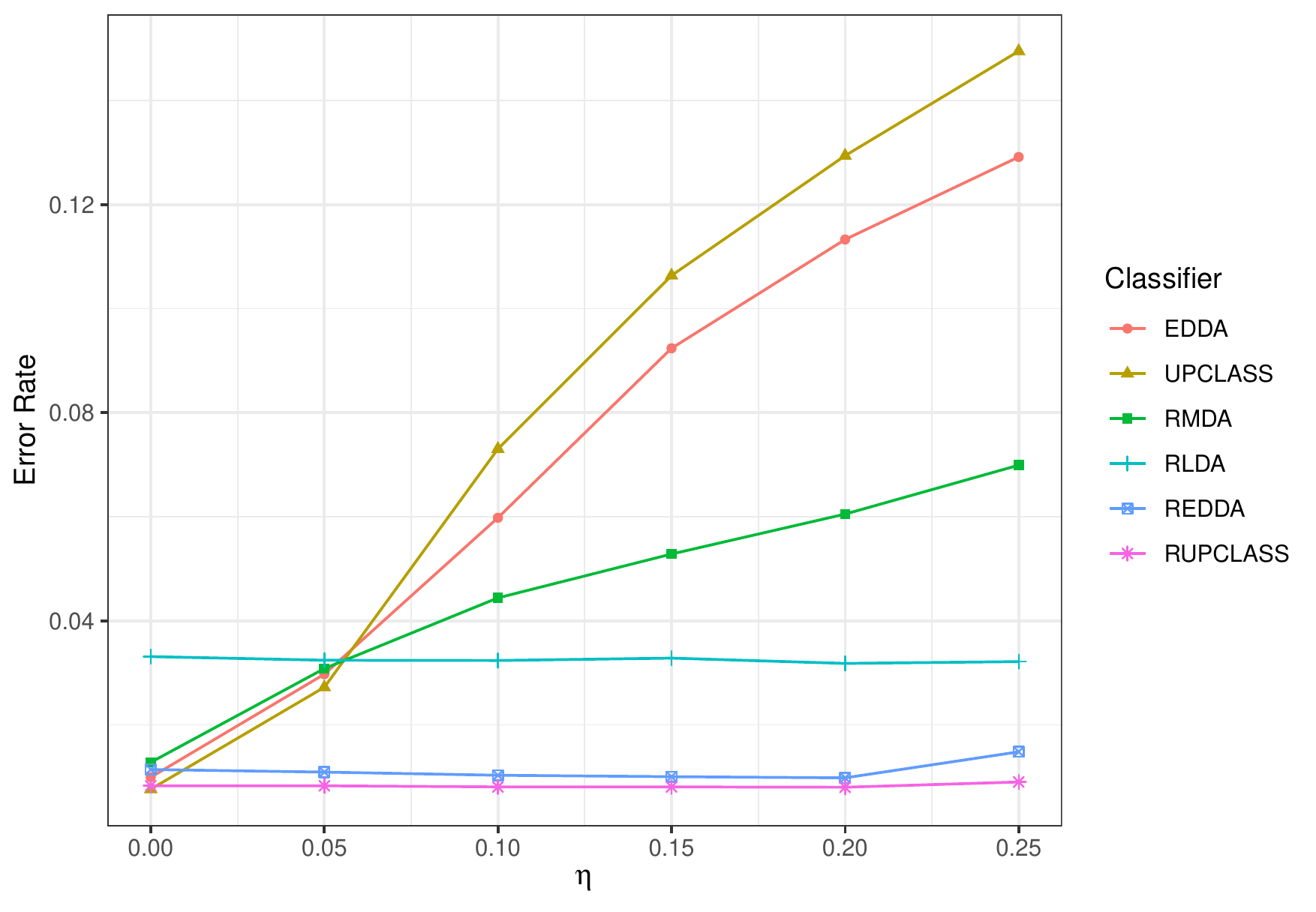}
% figure caption is below the figure
\caption{Average misclassification errors on $B=1000$ runs for different classification methods, increasing contamination rate.}
\label{fig:classerr}     % Give a unique label
\end{figure}
\section{Simulation studies} \label{sec:sim_study}
 
In this Section, we present two simulated data experiments: Simulation Study I compares the performances of several model-based classification methods in a low dimensional setting when dealing with noisy data at different contamination rates; Simulation Study II considers a higher dimensional scenario in which the accuracy performance of some popular classification methods is assessed, at a fixed contamination rate. In both scenarios we consider a joint noise structure on response and exploratory variables. %For each contamination rate and analysis, experiments have been repeated $B=1000$ times.}
\subsection{Simulation Study I}
\subsubsection{Experimental Setup} \label{sim_setup}
 We consider a data generating process given by a mixture of $G=3$ components of bivariate normal distributions, according to the following parameters:
 \[ \boldsymbol{\tau}=(0.3, 0.2, 0.5)', \quad \boldsymbol{\mu}_1=(0, 0)', \quad \boldsymbol{\mu}_2=(4, -4)', \quad \boldsymbol{\mu}_3=(0, 8)'\]
 \[ \boldsymbol{\Sigma}_1 = \begin{bmatrix}
    1       & 0.3\\
    0.3       & 1
    \end{bmatrix} \quad
    \boldsymbol{\Sigma}_2 = \begin{bmatrix}
    1       & -0.3\\
    -0.3       & 1
    \end{bmatrix} \quad
    \boldsymbol{\Sigma}_3 = \begin{bmatrix}
    6.71       & 2.09\\
    2.09       & 6.71
    \end{bmatrix}.
    \]
$600$ observations were generated from the model, randomly assigning $N=200$ to the labelled set and $M=400$ to the unlabelled set. The labelled set was subsequently adulterated with contamination rate $\eta$ (ranging from $0$ to $0.25$), %randomly switching $\lceil \eta N\rceil$ of the known labels
wrongly assigning $\lceil\eta/2 N\rceil$ of the third group units to the first class
and adding $\lceil \eta/2 N\rceil$ randomly labelled points generated from a Uniform distribution on the square with vertices $\left[(-20,-20), (-20,20), (20,-20), (20,20)\right]$. The contamination is therefore twofold, involving jointly label switching and outliers for a total of $\eta N$ adulterated labelled units. Examples of labelled datasets with different contamination rates are reported in Figure \ref{fig:cont}.
Performances of 6 model-based classification methods are considered:
%\begin{table}
%\caption{Average misclassification errors on $B=1000$ runs, varying method and contamination rate $\eta$. Standard errors are reported in parenthesis.}
%\label{tab:misclass}
%\centering
%%\begin{scriptsize}
%%\setlength{\tabcolsep}{4pt}
%\begin{tabular}{l|cccccc}
% \hline\noalign{\smallskip}
% \multicolumn{1}{r|}{$\eta$} & $0$  & $0.05$  & $0.10$  & $0.15$  & $0.20$  & $0.25$  \\ 
%\noalign{\smallskip}\hline\noalign{\smallskip}
%EDDA & 0.01 & 0.03 & 0.06 & 0.092 & 0.113 & 0.129 \\ 
%   & (0.005) & (0.026) & (0.049) & (0.06) & (0.059) & (0.059) \\ 
%  UPCLASS & 0.008 & 0.027 & 0.073 & 0.106 & 0.129 & 0.149 \\ 
%   & (0.004) & (0.04) & (0.076) & (0.081) & (0.076) & (0.065) \\ 
%  RMDA & 0.013 & 0.031 & 0.044 & 0.053 & 0.061 & 0.07 \\ 
%   & (0.021) & (0.038) & (0.039) & (0.043) & (0.043) & (0.048) \\ 
%  RLDA & 0.033 & 0.032 & 0.032 & 0.033 & 0.032 & 0.032 \\ 
%   & (0.013) & (0.013) & (0.013) & (0.013) & (0.013) & (0.013) \\ 
%  REDDA & 0.011 & 0.011 & 0.01 & 0.01 & 0.01 & 0.015 \\ 
%   & (0.005) & (0.005) & (0.005) & (0.005) & (0.005) & (0.01) \\ 
%  RUPCLASS & 0.008 & 0.008 & 0.008 & 0.008 & 0.008 & 0.009 \\ 
%   & (0.004) & (0.004) & (0.004) & (0.004) & (0.004) & (0.006) \\    
%  \noalign{\smallskip}\hline\noalign{\smallskip}
%\end{tabular}
%%\end{scriptsize}
%\end{table}

\begin{table}
\caption{Average misclassification errors on $B=1000$ runs, varying method and contamination rate $\eta$. Standard errors are reported in parenthesis.}
\label{tab:misclass}
\centering
\begin{tabular}{l|cccccc}
 \hline\noalign{\smallskip}
 \multicolumn{1}{r|}{$\eta$} & $0$  & $0.05$  & $0.10$  & $0.15$  & $0.20$  & $0.25$  \\ 
\noalign{\smallskip}\hline\noalign{\smallskip}
EDDA & 0.009 & 0.031 & 0.053 & 0.079 & 0.099 & 0.112 \\ 
 & (0.005) & (0.026) & (0.043) & (0.051) & (0.054) & (0.05) \\ 
 UPCLASS & 0.008 & 0.041 & 0.091 & 0.142 & 0.166 & 0.186 \\ 
  & (0.004) & (0.056) & (0.088) & (0.088) & (0.08) & (0.067) \\ 
RMDA &  0.009 & 0.045 & 0.049 & 0.052 & 0.07 & 0.08 \\ 
  & (0.005) & (0.072) & (0.063) & (0.057) & (0.068) & (0.073) \\ 
 RLDA &  0.027 & 0.027 & 0.026 & 0.026 & 0.026 & 0.067 \\ 
  & (0.009) & (0.009) & (0.009) & (0.008) & (0.009) & (0.037) \\ 
  REDDA &  0.01 & 0.01 & 0.01 & 0.009 & 0.024 & 0.042 \\ 
  & (0.005) & (0.005) & (0.005) & (0.005) & (0.014) & (0.014) \\ 
RUPCLASS &  0.01 & 0.009 & 0.009 & 0.008 & 0.019 & 0.044 \\ 
  & (0.005) & (0.005) & (0.005) & (0.005) & (0.013) & (0.014) \\ 
  \noalign{\smallskip}\hline\noalign{\smallskip}
\end{tabular}
%\end{scriptsize}
\end{table}

\begin{itemize}
\item EDDA: Eigenvalue Decomposition Discriminant Analysis \citep{Bensmail1996}
\item UPCLASS: Updating Classification Rules \citep{Dean2006}
\item RMDA: Robust Mixture Discriminant Analysis \citep{Bouveyron2009a}
\item RLDA: Robust Linear Discriminant Analysis \citep{Hawkins1997}
\item REDDA: Robust Eigenvalue Decomposition Discriminant Analysis. This is the supervised version of the model described in Section \ref{sec:REDDA}, where only the labelled observations are used for parameter estimation obtained via the robust initialization detailed in Section \ref{EM_alg}.% enforcing impartial trimming as in \eqref{cond_dens}
\item RUPCLASS: Robust Updating Classification Rules. The semi-supervised method described in Section \ref{sec:REDDA}.
\end{itemize}
To make a fair performance comparison, a level of $\alpha_l=0.15$ (REDDA and RUPCLASS) and $\alpha_u=0.05$ (RUPCLASS) have been kept fixed throughout the simulation study. Nevertheless, exploratory tools such as Density-Based Silhouette plot \citep{Menardi2011} and trimmed likelihood curves \citep{Garcia-Escudero2011} could be employed to validate and assess the choice of $\alpha_l$ and $\alpha_u$. A more automatic approach, like the one introduced in \cite{Dotto2018}, could also be adapted to our framework. This, however, goes beyond the scope of the present manuscript, it will nonetheless be addressed in the future. A value of $c=20$ was selected for the eigenvalue-ratio restriction in \eqref{eigen_contr}. Simulation study results are presented in the following subsections.

\subsubsection{Classification Performance}
\begin{figure}
 \centering 
% Use the relevant command to insert your figure file.
% For example, with the graphicx package use
% \includegraphics[keepaspectratio,width=\textwidth]{boxplot_param_2.pdf}
%\includegraphics[height=11cm,width=\textwidth]{sim_study_I_boxplot.pdf}
\includegraphics[width=\textwidth, keepaspectratio]{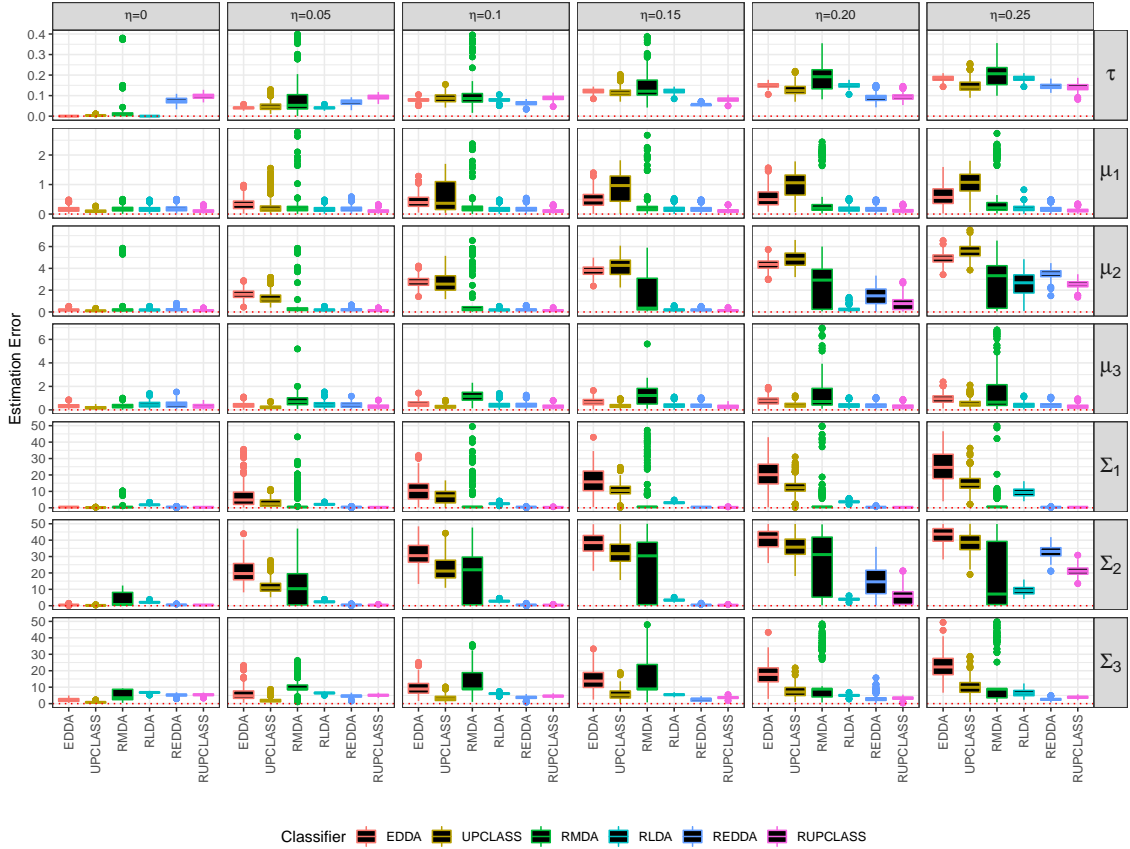}
% figure caption is below the figure
\caption{Box plots of the simulated estimation errors for the parameters of the mixture, computed via Euclidean norms for the proportion vector $\boldsymbol{\tau}$, the mean vectors $\boldsymbol{\mu}_g$ and covariance matrices $\boldsymbol{\Sigma}_g$, $g=1,\ldots,3$ for the different models, varying contamination rate $\eta$ from $0$ to $0.25$.}
\label{fig:boxparam}     % Give a unique label
\end{figure}

Average misclassification errors for the different methods and for varying contamination rates are reported in Table \ref{tab:misclass} and in Figure \ref{fig:classerr}. The error rate is computed on the unlabelled dataset and averaged over the $B=1000$ simulations. As expected, the misclassification error is fairly equal to all methods when there is no contamination rate, with the only exception being RLDA: this is due to the implicit model assumption that $\boldsymbol{\Sigma}_1=\boldsymbol{\Sigma}_2=\boldsymbol{\Sigma}_3$, which is not the case in our simulated scenario. As the contamination rate increases, so does the error rate for the non-robust methods (EDDA and UPCLASS), whereas for RLDA and RMDA it has a lower increment rate. Nevertheless, such methods fail to jointly cope with both sources of adulteration, namely class and attribute noise. Our proposals REDDA and RUPCLASS, thanks to the trimming step enforced in the estimation process, have always higher correct classification rates, on average, at any adulteration level. Notice that, to compare results of robust and non-robust methods, also the trimmed observations were classified a-posteriori according to the Bayes rule, assigning them to the component \(g\) having greater value
of $\hat{\tau}_g \phi(\mathbf{y}_m;\hat{\boldsymbol{\mu}}_g,\hat{\boldsymbol{\Sigma}}_g)$.

On average, the robust semi-supervised approach performs better than the supervised counterpart, due to the information incorporated from genuine unlabelled data in the estimation process. Interestingly, the same behavior is not reflected in the non-robust counterparts, where the detrimental effect of contaminated labelled units magnifies the bias of the UPCLASS method. Therefore, robust  solutions are even more paramount when a semi-supervised approach is considered.
% inclusion of the unlabeled data points (genuine and without added adulteration in this simulated scenario) in the estimation process.  
        
\subsubsection{Parameter Estimation}

%Table \ref{tab:bias_MSE} reports the average bias and MSE for the mixture parameters of the first component (computed element-wise for every element) for each method and adulteration rate. In a similar fashion, the box plots of the simulated distributions over $B=1000$ repetitions of the experiment are reported in Figure \ref{fig:boxparam}.
Figure \ref{fig:boxparam} reports the box plots of the simulated estimation error over $B=1000$ Monte Carlo repetitions for the parameters of the mixture model, computing Euclidean norms for the proportion vector $\boldsymbol{\tau}$, the mean vectors $\boldsymbol{\mu}_g$ and covariance matrices $\boldsymbol{\Sigma}_g$, $g=1,\ldots,3$.
The estimated values for the mixing proportion are mildly affected when increasing contamination is considered; conversely, the estimation of $\boldsymbol{\mu}_2$ is on average heavily influenced by the adulterating process, and also the robust methods fail to estimate it correctly as soon as the contamination rate $\eta$ is larger than the trimming level $\alpha_l=0.15$. Clearly, the estimation of the variance covariance matrices is as well badly affected in most extreme scenarios, where their entries are inflated in order to accommodate more and more bad points. Our robust proposals are less affected by the harmful effect of adding anomalous observations, also in the most adulterated scenario.

\subsection{Simulation Study II}
\subsubsection{Experimental Setup} \label{sim_setup_2}
We consider here a simulating model with a larger number of features ($d=10$), where the data generating process is given by a mixture of $G=4$ components of a multivariate t-distribution with $\nu=6$ degrees of freedom. More details on the parameter values are contained in Appendix B. $1000$ observations were generated from the model, randomly assigning $N=250$ to the labelled set and $M=750$ to the unlabelled set. The training set was subsequently adulterated wrongly labelling $10$ units and adding $15$ randomly labelled outlying points, uniformly generated in the  $d$-dimensional hypercube over $[10,15]^{10}$. We therefore consider a scenario in which $10\%$ of the learning units are contaminated, via both label and attribute noise.

Together with the model-based methods previously described in Section \ref{sim_setup}, we included in the performance evaluation widely used classification techniques that, even though not engineered to achieve robustness, are noise tolerant. Particularly, the ensemble learner AdaBoost \citep{Schapire1997a} and the kernel method Support Vector Machine \citep{Farhat2002} were added to the comparison. Furthermore, the robust adaptation of the SIMCA method for high-dimensional classification \citep{VandenBranden2005} was also considered. The classification performance of the afore-described techniques are tested against the proposed methodologies, under different combinations of $(c, \alpha_l, \alpha_u)$: accuracy results are reported in the next Section.

\begin{figure}
 \centering 
% Use the relevant command to insert your figure file.
% For example, with the graphicx package use
% \includegraphics[keepaspectratio,width=\textwidth]{boxplot_param_2.pdf}
\includegraphics[width=\textwidth, keepaspectratio]{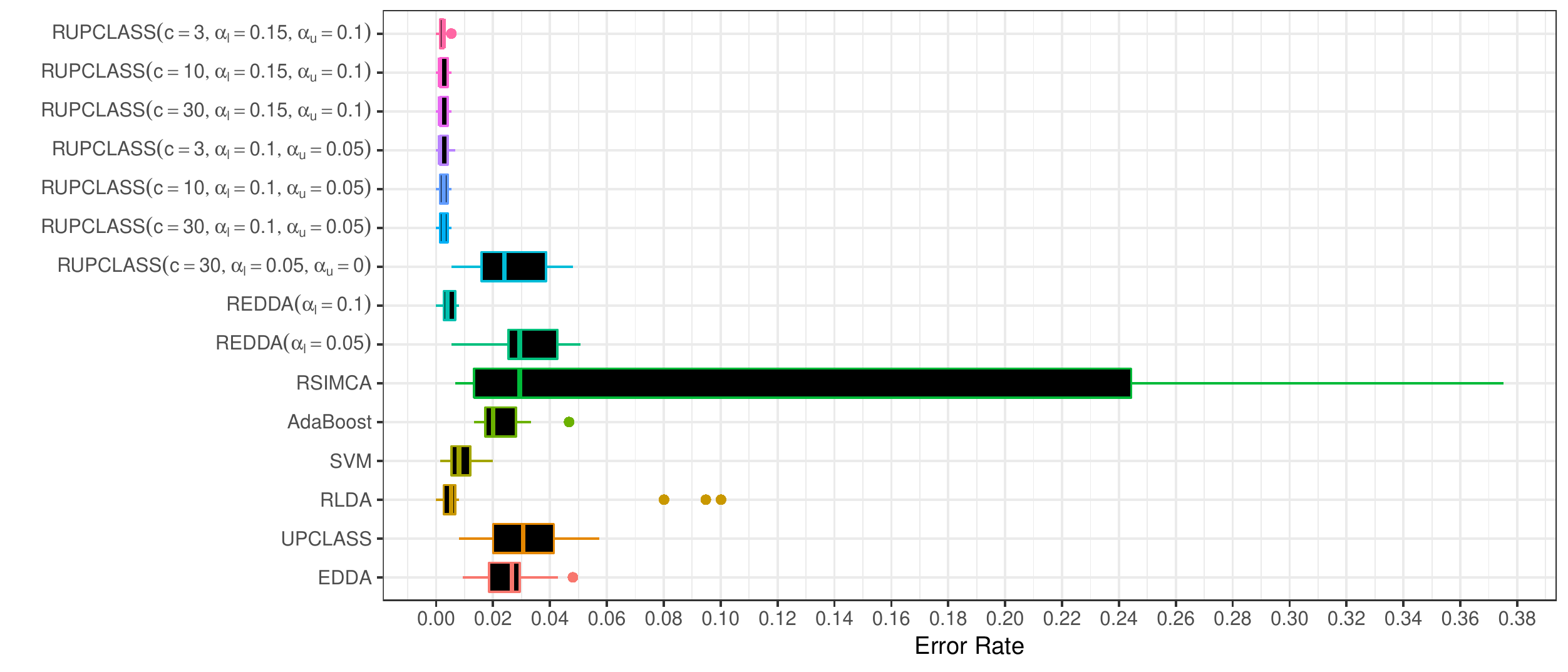}
% figure caption is below the figure
\caption{Box plots of the misclassification errors under $B=1000$ repetitions of the simulating experiment II. Error rate is computed on the $M=750$ data points of the test set for different classification methods.}
\label{fig:box_misclass}     % Give a unique label
\end{figure}
\subsubsection{Classification Performance}
Boxplots of the misclassification errors for the considered methods are reported in Figure \ref{fig:box_misclass}. The error rate is computed on the $M=750$ units of the test set, under $B=1000$ repetitions of the generating process and subsequent adulteration scheme described in Section \ref{sim_setup_2}. As it was already apparent from the previous simulation study, accuracy for non-robust methods is badly affected by the contamination present in the learning set. Even though not specifically designed for dealing with adulterated datasets, SVM and AdaBoost perform better than the non-robust model-based approaches, thanks to their non-parametric nature and flexibility. As expected, the best classification accuracy are obtained by the robust methodologies, namely RLDA and our proposals REDDA and RUPCLASS. We also check the sensitivity of our techniques comparing different combinations of $(c, \alpha_l, \alpha_u)$. As it is easily visible in the boxplots, setting a smaller than needed labelled trimming level $\alpha_l$ leads to a  loss in prediction accuracy, as a portion of adulterated units still affects the learning phase. Once the corrupted observations are correctly trimmed (i.e., $\alpha_l \text{ is set} \geq 0.1$), accuracy seems to remain stable with little influence induced by the choice of $c$ and $\alpha_u$, with only a slight preference for the semi-supervised RUPCLASS over its supervised version REDDA. This shows that setting a higher value of $\alpha_l$ is less detrimental than underestimating it, and that the impartial trimming almost exactly identifies the corrupted units when $\alpha_l=0.1$, that is the true adulteration proportion. The bad performance of RSIMCA is only due to the simulating process: given the fact that data truly lie on a 10-dimensional space, performing (robust) dimensional reduction prior to classification evidently leads to a concealment in the grouping structure.

The proposed methodologies were shown to be capable of dealing with data whose distribution is not exactly Gaussian, but where an effective robust decision rule can be built employing Gaussian mixture models.

\section{Application to Midinfrared Spectroscopy of Irish Honey} \label{sec:Honey}
The semi-supervised method introduced in Section \ref{sec:REDDA} is employed in performing adulteration detection and classification in  a food authenticity context: we consider the task of discriminating between pure and adulterated Irish Honey, where the training set itself contains unreliable samples.
\begin{figure}
 \centering 
% Use the relevant command to insert your figure file.
% For example, with the graphicx package use
 \includegraphics[height=3.4cm, width=\textwidth]{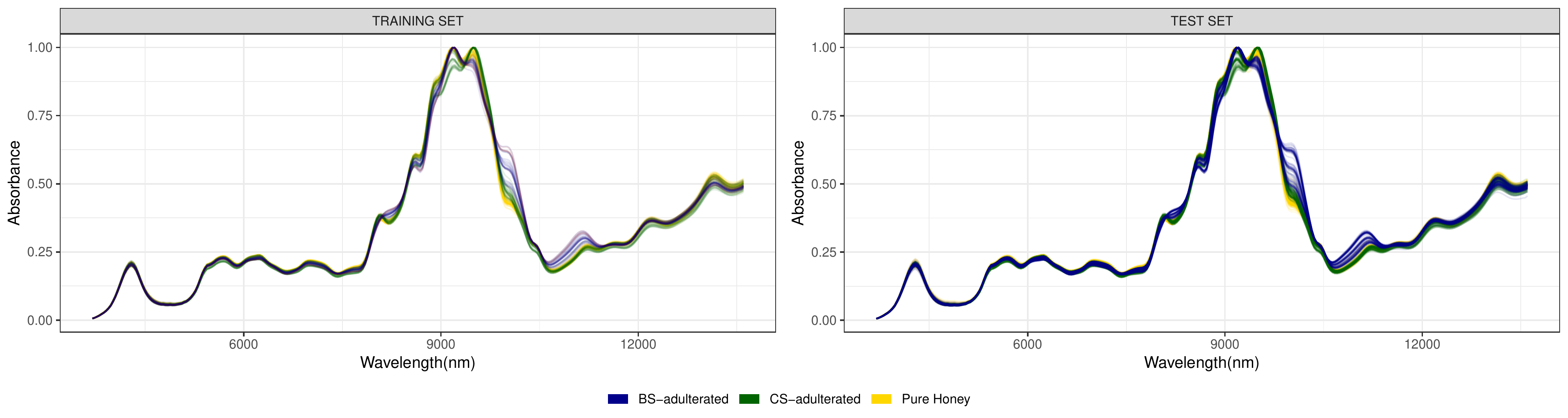}
% figure caption is below the figure
\caption{Midinfrared spectra for pure and contaminated honey, Irish Honey data.} %Red spectra denotes adulterated samples wrongly labelled as Pure Honey in the Training Set.}
\label{fig:honey}     % Give a unique label
\end{figure}

\subsection{Honey Samples}
Honey is defined as ``the natural sweet substance, produced by honeybees from the nectar of plants or from secretions of living parts of plants, or excretions of plant-sucking insects on the living parts of plants, which the bees collect, transform by combining with specific substances of their own, deposit, dehydrate, store and leave in honeycombs to ripen and mature'' \citep{alimentarius2001revised}. Being a relatively expensive commodity to produce and extremely variable in nature, honey is prone to adulteration for economic gain: in 2015 the European Commission organized an EU coordinated control plan to assess the prevalence on the market of honey adulterated with sugars and honeys mislabelled with regard to their botanical source or geographical origin. It is therefore of prime interest to employ robust analytical methods to protect food quality and uncover its illegal adulteration.

We consider here a dataset of midinfrared spectroscopic measurements of 530 Irish honey samples. Midinfrared spectroscopy is a fast, non-invasive method for examining substances that does not require any sample preparation, it is therefore an effective procedure for collecting data to be subsequently used in food authenticity studies \citep{doi:10.1255/jnirs.75}. The spectra measurements lie in the wavelength range of $3700~\mbox{nm}$ and $13600~\mbox{nm}$, recorded at intervals of $35~\mbox{nm}$, with a total of 285 absorbance values. The dataset contains 290 Pure Honey observations, while the rest of the samples are honey diluted with adulterant solutions: 120 with Dextrose Syrup and 120 with Beet Sucrose, respectively. \cite{Kelly2006} gives a thorough explanation of the adulteration process. The aim of the study is to discriminate pure honey from the adulterated samples, when varying sample size of the labelled set whilst including a percentage of wrongly labelled units. Such a scenario is plausible to be encountered in real situations, since in a context in which the final purpose is to detect potential adulterated samples it may happen that the learning data is itself not fully reliable. An example of the data structure is reported in Figure \ref{fig:honey}.
\begin{figure}
 \centering 
% Use the relevant command to insert your figure file.
% For example, with the graphicx package use
 \includegraphics[width=\textwidth, height=3cm]{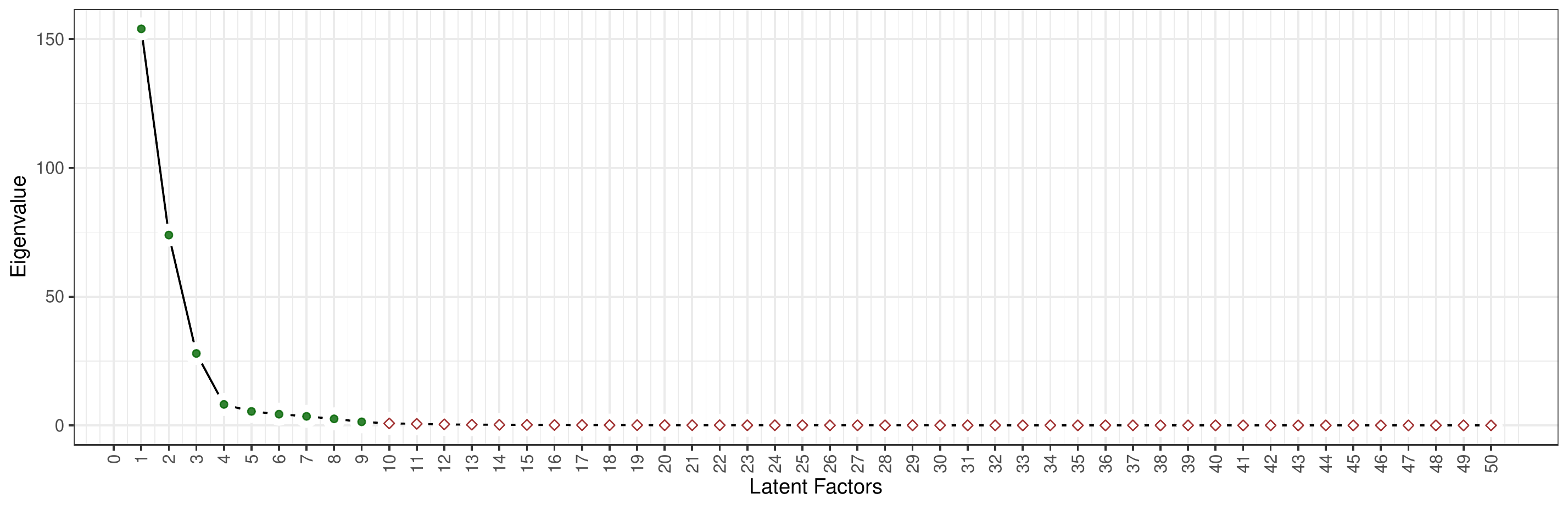}
% figure caption is below the figure
\caption{Cattell's scree plot \citep{Cattell1966} for the first 50 eigenvalues of the robustly estimated correlation matrix, Irish Honey data. Green solid dots denote eigenvalues bigger than 1.}
\label{fig:scree_p}     % Give a unique label
\end{figure}            
\subsection{Robust Dimensional Reduction} \label{rMFA}

Prior to perform classification and adulteration detection, a preprocessing step is needed due to the high-dimensional nature of the considered dataset ($p=285$ variables). To do so, we robustly estimate a factor analysis model, retaining a set $d$ of factors, $d \ll p$, to be subsequently employed with the Robust Updating Classification Rules. Formally, for each Honey sample $\mathbf{x}_i$, we postulate a factor model of the form:
\begin{equation} \label{FA}
\mathbf{x}_i= \boldsymbol{\mu} + \boldsymbol{\Lambda}\boldsymbol{u}_i+\boldsymbol{e}_i
\end{equation}
where $\boldsymbol{\mu}$ is a $p \times 1$ mean vector, $\boldsymbol{\Lambda}$ is a $p\times d$ matrix of factor loadings, $\boldsymbol{u}_i$ are the unobserved factors, assumed to be realizations of a $d$-variate standard normal and the errors $\boldsymbol{e}_i$ are independent realizations of $\mathcal{N}(\mathbf{0}, \boldsymbol{\Psi})$, with $\boldsymbol{\Psi}$ a $p \times p$ diagonal matrix. In such a way, the observed variables are assumed independent given the factors. For a general review on factor analysis, see for example Chapter 9 in \citep{bibby1979multivariate}.
Parameters in \eqref{FA} are estimated employing a robust procedure based on trimming and constraints \citep{garcia2016joint}, yielding dimensionality reduction at the same time. Given the  robustly estimated parameters, the latent traits are computed using the regression method \citep{thomson1946factorial}:
\begin{equation}
\hat{\boldsymbol{u}}_i=\hat{ \boldsymbol{\Lambda}}^{'}\left(\hat{ \boldsymbol{\Lambda}}\hat{\boldsymbol{\Lambda}}^{'}+\hat{\boldsymbol{\Psi}}\right)^{-1}(\mathbf{x}_i-\bar{\mathbf{x}})
\end{equation}
The estimated factors scores $\hat{\boldsymbol{u}}_i$ will be used for the classification task reported in the upcoming Section. %For the considered dataset, after a careful investigation (not reported here) we deem sufficient to set the number $d$ of latent factors equal to 10, with trimming level $\alpha=0.1$ and $c_{noise}=1000$.
For the considered dataset, after a graphical exploration of Cattell's scree plot for the correlation matrix robustly estimated via MCD \citep{Driessen1999}, reported in Figure \ref{fig:scree_p}, we deem sufficient to set the number $d$ of latent factors equal to $10$. Parameters were estimated setting a trimming level $\alpha=0.1$ and $c_{noise}=1000$.  
\subsection{Classification Performance}
SVM, AdaBoost and RSIMCA are designed to optimally perform in a high dimensional setting. Therefore, to respect the specificity of each family of methodologies, we directly applied SVM, AdaBoost and RSIMCA on the whole spectra. For EDDA, UPCLASS, RMDA, RLDA, REDDA and RUPCLASS we preprocessed the data with the dimension reduction method described in Section \ref{rMFA}. To discriminate between pure and adulterated honey samples, we divided the available data into a training (labelled) sample and a validation (unlabelled) sample. We investigated the effect of having different sample sizes in the labelled set, both in terms of classification accuracy and adulteration detection. Particularly, 3 proportions have been considered: \(50\%\) - \(50\%\) ,
%After having performed robust dimensional reduction, the method described in Section \ref{sec:REDDA} \textcolor{blue}{and all the classification techniques included in the simulation studies} have been employed for discriminating between pure and adulterated honey samples. To do so, we divided the available data into a training (labelled) sample and a validation (unlabelled) sample. We investigated the effect of having different sample sizes in the labelled set, both in terms of classification accuracy and adulteration detection. Particularly, 3 proportions have been considered: \(50\%\) - \(50\%\) ,
\(25\%\) - \(75\%\)  and \(10\%\) - \(90\%\)
for splitting data into training and validation set, respectively, within each group. For each split, $10\%$ of the Beet Sucrose adulterated samples were incorrectly labelled as Pure Honey in the training set, adding class noise in the discrimination task. The trimming levels $\alpha_l$ and $\alpha_u$ were set equal to $0.12$ and $0.05$, respectively. Table \ref{tab:honey_2} and \ref{tab:honey} summarize the accuracy results employing different classification approaches under the described scenarios. Careful investigation has been dedicated to measuring the ability of the robust methodologies in correctly determining (i.e., trimming) the $10\%$ of incorrectly labelled samples, that is, units adulterated with Beet Sucrose and erroneously labelled as Pure Honey: such information, only relevant for RSIMCA, REDDA and RUPCLASS models, is reported in Table \ref{tab:honey}. \textit{\% Correctly Trimmed} indicates the class noise percentage correctly detected by the impartial trimming. For the recognized class noise, \textit{\% Correctly Assigned} indicates the percentage of units properly a-posteriori assigned to the Beet Sucrose group. RSIMCA performs remarkably well in identifying the adulterated units, even though the classification accuracy is lower than the one obtained employing RUPCLASS model.  
As expected, the semi-supervised approach performs much better in terms of classification rate when the labelled sample size is small. Comparing the error rate of the robust techniques with the other methods in Table \ref{tab:honey_2} we notice how powerful classifiers like SVM and AdaBoost work well also in dealing with adulterated datasets: SVM error rate in the 50\% Tr - 50\% Te is on average lower than the one obtained with RUPCLASS. However, when the labelled sample size decreases a semi-supervised approach is preferable: RUPCLASS reports the lowest error rate for both 25\% Tr - 75\% Te and 10\% Tr - 90\% Te scenarios.
VEV and VVV models have been almost always chosen:
model selection was performed through the Robust criteria defined in Section \ref{sec:RBIC}. %\textcolor{blue}{As a last worthy comment, notice that for methods that automatically deal with high-dimensional data (i.e., SVM, AdaBoost and RSIMCA) no dimension reduction was performed, hence proving the efficacy of the technique described in Section \ref{rMFA}.} 

Results in Table \ref{tab:honey} show that the proposed methodology is effective not only for accurately robustifying the parameter estimates%that translates into an increase in the classification rate
, but also for efficiently detecting observations affected by class noise, firstly by trimming and subsequently by correctly assigning them: a critical information that cannot be obtained with standard classification methods like SVM and AdaBoost.
 \begin{table}
\centering
\caption{Misclassification rates in the unlabelled set for different classification methods.  
Average values for 50 random splits in training and validation (three proportions are considered), standard deviations reported in
parentheses.}
\label{tab:honey_2}
\begin{tabular}{l|llllll}
  \noalign{\smallskip}\hline\noalign{\smallskip}
Error Rate & EDDA & UPCLASS & RMDA & RLDA & SVM & AdaBoost \\
 \noalign{\smallskip}\hline\noalign{\smallskip}
50\% Tr - 50\% Te & 0.033 & 0.065 & 0.291 & 0.1 & 0.025 & 0.036 \\
 & (0.012) & (0.049) & (0.091) & (0.02) & (0.008) & (0.011) \\
  25\% Tr - 75\% Te & 0.078 & 0.112 & 0.303 & 0.12 & 0.048 & 0.042 \\
 & (0.025) & (0.028) & (0.08) & (0.04) & (0.021) & (0.012) \\
 10\% Tr - 90\% Te & 0.24 & 0.126 & 0.375 & 0.157 & 0.109 & 0.058 \\
   & (0.031) & (0.023) & (0.065) & (0.08) & (0.036) & (0.021) \\
   \noalign{\smallskip}\hline\noalign{\smallskip}
\end{tabular}
\end{table}   

\begin{table}
\centering
\caption{Misclassification rates in the unlabelled set, \% of wrongly labelled samples correctly trimmed in the labelled set and \% of those correctly trimmed observations properly a-posteriori assigned to the Beet Sucrose group.  
Average values for 50 random splits in training and validation (three proportions are considered), standard deviations reported in
parentheses.}
\label{tab:honey}
\begin{tabular}{l|llll}
  \noalign{\smallskip}\hline\noalign{\smallskip}
& &RSIMCA & REDDA & RUPCLASS \\ 
 \noalign{\smallskip}\hline\noalign{\smallskip}
50\% Tr - 50\% Te& Error Rate & 0.069 & 0.05 & 0.029 \\ 
& & (0.029) & (0.013) & (0.01) \\ 
  & \% Correctly Trimmed & 1 & 0.977 & 1 \\ 
&  & (0) & (0.075) & (0) \\ 
  & \% Correctly Assigned &1 & 1 & 1 \\ 
  & & (0)&(0) & (0) \\ \\
  25\% Tr - 75\% Te & Error Rate & 0.075 & 0.053 & 0.032 \\ 
& & (0.038) & (0.034) & (0.009) \\
  & \% Correctly Trimmed & 1 & 0.88 & 0.96 \\ 
&  & (0) & (0.25) & (0.145) \\ 
  & \% Correctly Assigned &1 & 0.963 & 1 \\ 
  & & (0)& (0.162) & (0) \\  \\
 10\% Tr - 90\% Te & Error Rate & 0.111 & 0.121 & 0.053 \\ 
  & & (0.051) & (0.039) & (0.038) \\ 
  &\% Correctly Trimmed  & 0.99 & 0.47 & 0.73 \\ 
   & & (0.071) & (0.238) & (0.381) \\ 
  & \% Correctly Assigned & 0.99& 0.72 & 0.84 \\ 
  & & (0.019) & (0.071) & (0.37) \\ 
   \noalign{\smallskip}\hline\noalign{\smallskip}
\end{tabular}
\end{table}

\section{Concluding Remarks} \label{sec:conclusion}

In this paper we have proposed a robust modification to a family of semi-supervised patterned models, for performing classification in presence of both class and attribute noise.

We have shown that our methodology effectively addresses the issues generated by these two noise types, by identifying wrongly labelled units (noise in the response variable) and corrupted attributes in units (noise in the explanatory variables). Robust parameter estimates can therefore be obtained by excluding the noisy observations from the estimation procedure, both in the training set, and in the test set. Our proposal has been based on incorporating  impartial trimming and eigenvalue-ratio constraints in previous semi-supervised methods. We have adapted the trimming procedure to the two different frameworks, i.e., for the labelled units and the unlabelled ones. 
After completing the robust estimation process, trimmed observations can be classified as well, by the usual Bayes rule. This final step allows the researcher to detect whether one observation is indeed extreme in terms of its attributes or it has been wrongly assigned to a different class. Such feature seems particularly desirable in food authenticity applications, where, due to imprecise readings and fraudulent units, it is likely to have label noise also within the labelled set. Some simulations, and a study on real data from pure and adulterated Honey samples, have shown the effectiveness of our proposal.

As an open point for further research, an automatic procedure for selecting reasonable  values for the labelled and unlabelled trimming levels, along the lines of \cite{Dotto2018}, is under study. Additionally, a robust wrapper variable selection for dealing with high-dimensional problems could be useful for further enhancing the discriminating power of the proposed methodology.
\vspace*{-.3cm}
%In this paper we have proposed a robust modification to a family of semi-supervised patterned models, for performing classification in presence of both class and attribute noise employing impartial trimming and eigenvalue-ratio constraints. Our proposal jointly accounts for these two noise types, producing robust parameter estimates excluding the noisy observations from the estimation procedure. Furthermore, a trimmed observation can then be classified a posteriori, allowing the researcher to detect whether the observation is indeed extreme in terms of its attributes or was wrongly assigned to a different class. Such asset seems particularly desirable in food authenticity applications, where it is likely to have fraudulent units also within the labelled set. %Applications to simulated and real datasets highlight good classification rates, 	 

%As an open point for further research, an automatic procedure for selecting a reasonable range of values for the labelled and unlabelled trimming levels, along the lines of \cite{Dotto2018}, seem desirable. Additionally, a robust wrapper variable selection for dealing with high-dimensional problems could be useful for further enhancing the discriminating power of the proposed robust model for data classification.
\subsubsection*{Acknowledgements}
The authors are very grateful to Agustin Mayo-Iscar and Luis Angel Garc\'{i}a Escudero for both stimulating discussion and advices on how to enforce the eigenvalue-ratio constraints under the different patterned models. Andrea Cappozzo deeply thanks Michael Fop for his endless patience and guidance in helping him with methodological and computational issues encountered during the draft of the present manuscript. Brendan Murphy's work is supported by the Science Foundation Ireland Insight Research Centre (SFI/12/RC/2289\_P2)

\newpage
%\vspace*{-.8cm} 
\section*{Appendix A}
\textbf{Proof of Proposition 1:} Considering the random variable $\mathcal{Z}_{mg}$  corresponding to $z_{mg}$, the E-step on the $(k+1)$th iteration requires the calculation of the conditional expectation of  $\mathcal{Z}_{mg}$ given $\mathbf{y}_m$:
\begin{align}
\begin{split}
\mathit{E}_{\hat{\boldsymbol{\theta}}^{(k)}}(\mathcal{Z}_{mg}|\mathbf{y}_m)&=\mathbb{P}\left(\mathcal{Z}_{mg}=1|\mathbf{y}_m;\hat{\theta}^{(k)}\right)=\\
&=\frac{\mathbb{P}\left(\mathbf{y}_m|\mathcal{Z}_{mg}=1;\hat{\theta}^{(k)}\right)\mathbb{P}\left(\mathcal{Z}_{mg}=1;\hat{\theta}^{(k)}\right)}{\sum_{j=1}^G \mathbb{P}\left(\mathbf{y}_m|\mathcal{Z}_{mj}=1;\hat{\theta}^{(k)}\right)\mathbb{P}\left(\mathcal{Z}_{mj}=1;\hat{\theta}^{(k)}\right)}=\\
&=\frac{\hat{\tau}^{(k)}_g \phi \left(\mathbf{y}_m; \hat{\boldsymbol{\mu}}^{(k)}_g, \hat{\boldsymbol{\Sigma}}^{(k)}_g \right)}{\sum_{j=1}^G\hat{\tau}_j^{(k)} \phi\left(\mathbf{y}_m; \hat{\boldsymbol{\mu}}^{(k)}_j, \hat{\boldsymbol{\Sigma}}^{(k)}_j\right)}=\\
&=\hat{z}_{mg}^{(k+1)}  \:\:\:\:\:\:\:\:\:\:\:\:\:\:\:\:\:\:\:\:\:\:\:\:\:\:\:\:\:\:\:\:\:\:\: g=1,\ldots, G; \:\:\:\: m=1,\ldots, M.
\end{split}
\end{align}
Therefore, the Q function, to be maximized with respect to $\boldsymbol{\theta}$ in the M-step, is given by
\begin{align} \label{Q_M_step}
\begin{split}
Q(\boldsymbol{\theta};\hat{\boldsymbol{\theta}}^{(k)})&=
\sum_{n=1}^N \zeta(\mathbf{x}_n) \sum_{g=1}^G l_{ng} \log{\left[\tau_g \phi(\mathbf{x}_n; \boldsymbol{\mu}_g, \boldsymbol{\Sigma}_g)\right]} +\\
&+ \sum_{m=1}^M \varphi(\mathbf{y}_m) \sum_{g=1}^G \hat{z}_{mg} \log{\left[\tau_g \phi(\mathbf{y}_m; \boldsymbol{\mu}_g, \boldsymbol{\Sigma}_g)\right].}
\end{split}
\end{align}
The maximization of \eqref{Q_M_step} according to the mixture proportion $\tau_g$, $\sum_{j=1}^G\tau_j=1$ is solved considering the Lagrangian $\mathcal{L}(\boldsymbol{\theta}, \kappa)$:
\begin{equation} \label{lagran}
\mathcal{L}(\boldsymbol{\theta}, \kappa)=Q(\boldsymbol{\theta};\hat{\boldsymbol{\theta}}^{(k)})-\kappa\left(\sum_{j=1}^G\tau_j-1\right)
\end{equation}
with $\kappa$ the Lagrangian coefficient. The partial derivative of \eqref{lagran} with respect to $\tau_g$ has the form:
\begin{equation}\label{deriv_tau}
\frac{\partial}{\partial\tau_g}\mathcal{L}(\boldsymbol{\theta}, \kappa)=\frac{\sum_{n=1}^N \zeta(\mathbf{x}_n)l_{ng}}{\tau_g}+ \frac{\sum_{m=1}^M \varphi(\mathbf{y}_m)\hat{z}_{mg}}{\tau_g}-\kappa
\end{equation}
and setting \eqref{deriv_tau} equal to $0$ for all $g=1,\ldots, G$ we obtain:
\begin{equation}\label{deriv_tau_0}
\sum_{n=1}^N \zeta(\mathbf{x}_n)l_{ng}+ \sum_{m=1}^M \varphi(\mathbf{y}_m)\hat{z}_{mg}-\kappa\tau_g=0.
\end{equation}
Summing \eqref{deriv_tau_0} over $g$, $g=1,\ldots, G$, provides the value of $\kappa=\lceil N(1-\alpha_{l})\rceil+M(1-\alpha_{u})\rceil$ and substituting it in the previous expression yields the ML estimate for $\tau_g$:
\begin{equation}
\hat{\tau}_g^{(k+1)}=\frac{\sum_{n=1}^N \zeta(\mathbf{x}_n)l_{ng}+ \sum_{m=1}^M \varphi(\mathbf{y}_m)\hat{z}_{mg}^{(k+1)}}{\lceil N(1-\alpha_{l})\rceil+\lceil M(1-\alpha_{u})\rceil}\:\:\:\:\: g=1,\ldots, G.\\
\end{equation}
The partial derivative of \eqref{Q_M_step} with respect to the mean vector $\boldsymbol{\mu}_g$ reads:
\begin{align} \label{deriv_mu}
\begin{split}
\frac{\partial}{\partial \boldsymbol{\mu}_g}Q(\boldsymbol{\theta};\boldsymbol{\theta}^{(k)})&=
-\boldsymbol{\Sigma}_g^{-1}\left[\sum_{n=1}^N \zeta(\mathbf{x}_n)l_{ng}\left(\mathbf{x}_n-\boldsymbol{\mu}_g\right)+\sum_{m=1}^M \varphi(\mathbf{y}_m)\hat{z}_{mg}^{(k+1)}\left(\mathbf{y}_m-\boldsymbol{\mu}_g\right)\right]=\\
&=-\boldsymbol{\Sigma}_g^{-1}\Bigg[\sum_{n=1}^N \zeta(\mathbf{x}_n)l_{ng}\mathbf{x}_n + \sum_{m=1}^M \varphi(\mathbf{y}_m)\hat{z}_{mg}^{(k+1)}\mathbf{y}_m +\\
&-\boldsymbol{\mu}_g\left( \sum_{n=1}^N \zeta(\mathbf{x}_n)l_{ng} + \sum_{m=1}^M \varphi(\mathbf{y}_m)\hat{z}_{mg}^{(k+1)} \right)\Bigg].
\end{split}
\end{align}
Equating \eqref{deriv_mu} to $0$ and rearranging terms provides the ML estimate of $\boldsymbol{\mu}_g$:
\begin{equation}
\hat{\boldsymbol{\mu}}_g^{(k+1)}=\frac{\sum_{n=1}^N \zeta(\mathbf{x}_n)l_{ng}\mathbf{x}_n+\sum_{m=1}^M\varphi(\mathbf{y}_m)\hat{z}_{mg}^{(k+1)}\mathbf{y}_m}{\sum_{n=1}^N\zeta(\mathbf{x}_n)l_{ng}+\sum_{m=1}^M\varphi(\mathbf{y}_m)\hat{z}_{mg}^{(k+1)}}\:\:\:\:\: g=1,\ldots, G.
\end{equation}
Discarding quantities that do not depend on $\boldsymbol{\Sigma}_g$, we can rewrite \eqref{Q_M_step} as follows:

\begin{multline} \label{q_mstep_sigma}
\sum_{n=1}^{N} \sum_{g=1}^{G} \zeta(\mathbf{x}_n)l_{ng}\left(\mathbf{x}_{n}\right)\left[-\log \left|\boldsymbol{\Sigma}_{g}\right|^{1 / 2}-\frac{1}{2}\left(\mathbf{x}_{n}-\boldsymbol{\mu}_{g}\right)^{\prime} \boldsymbol{\Sigma}_{g}^{-1}\left(\mathbf{x}_{n}-\boldsymbol{\mu}_{g}\right)\right]+\\
+\sum_{m=1}^{M} \sum_{g=1}^{G} \varphi(\mathbf{y}_m)\hat{z}_{mg}\left(\mathbf{y}_{m}\right)\left[-\log \left|\boldsymbol{\Sigma}_{g}\right|^{1 / 2}-\frac{1}{2}\left(\mathbf{y}_{m}-\boldsymbol{\mu}_{g}\right)^{\prime} \boldsymbol{\Sigma}_{g}^{-1}\left(\mathbf{y}_{m}-\boldsymbol{\mu}_{g}\right)\right]=\\
=-\frac{1}{2}\left[\sum_{n=1}^{N} \sum_{g=1}^{G} \zeta(\mathbf{x}_n)l_{ng}\left(\mathbf{x}_{n}\right) \log \left|\boldsymbol{\Sigma}_{g}\right|+\sum_{n=1}^{N} \sum_{g=1}^{G} \zeta(\mathbf{x}_n)l_{ng} \left[\underbrace{\left(\mathbf{x}_{n}-\boldsymbol{\mu}_{g}\right)^{\prime} \boldsymbol{\Sigma}_{g}^{-1}\left(\mathbf{x}_{n}-\boldsymbol{\mu}_{g}\right)}_{\text {a scalar }}\right]\right.+\\
+\left.\sum_{m=1}^{M} \sum_{g=1}^{G} \varphi(\mathbf{y}_m)\hat{z}_{mg}\left(\mathbf{y}_{m}\right) \log \left|\boldsymbol{\Sigma}_{g}\right|+\sum_{m=1}^{M} \sum_{g=1}^{G} \varphi(\mathbf{y}_m)\hat{z}_{mg} \left[\underbrace{\left(\mathbf{y}_{m}-\boldsymbol{\mu}_{g}\right)^{\prime} \boldsymbol{\Sigma}_{g}^{-1}\left(\mathbf{y}_{m}-\boldsymbol{\mu}_{g}\right)}_{ \text {a scalar }}\right]\right]=\\
=-\frac{1}{2}\left[ \sum_{g=1}^{G} \log \left|\boldsymbol{\Sigma}_{g}\right| \left(\sum_{n=1}^{N} \zeta(\mathbf{x}_n)l_{ng}\left(\mathbf{x}_{n}\right)+ \sum_{m=1}^{M} \varphi(\mathbf{y}_m)\hat{z}_{mg}\left(\mathbf{y}_{m}\right)\right)\right.+\\
+\sum_{n=1}^{N} \sum_{g=1}^{G} \zeta(\mathbf{x}_n)l_{ng} tr \left[\boldsymbol{\Sigma}_{g}^{-1} \left(\mathbf{x}_{n}-\boldsymbol{\mu}_{g}\right)\left(\mathbf{x}_{n}-\boldsymbol{\mu}_{g}\right)^{\prime}\right]+\\
\left.+ \sum_{m=1}^{M} \sum_{g=1}^{G} \varphi(\mathbf{y}_m)\hat{z}_{mg}tr \left[\boldsymbol{\Sigma}_{g}^{-1} \left(\mathbf{y}_{m}-\boldsymbol{\mu}_{g}\right)\left(\mathbf{y}_{m}-\boldsymbol{\mu}_{g}\right)^{\prime}\right]\right]=\\
=-\frac{1}{2}\left[ \sum_{g=1}^{G} \log \left|\boldsymbol{\Sigma}_{g}\right| \left(\sum_{n=1}^{N} \zeta(\mathbf{x}_n)l_{ng}\left(\mathbf{x}_{n}\right)+ \sum_{m=1}^{M} \varphi(\mathbf{y}_m)\hat{z}_{mg}\left(\mathbf{y}_{m}\right)\right)\right.+\\
+\left. \sum_{g=1}^{G} tr \left[\boldsymbol{\Sigma}^{-1}_{g}\boldsymbol{W}_g^{X} \right]+ \sum_{g=1}^{G} tr \left[\boldsymbol{\Sigma}_{g}^{-1}\boldsymbol{W}_g^{Y} \right] \right] =\\
= -\frac{1}{2}\left[ \sum_{g=1}^{G} \log \left|\boldsymbol{\Sigma}_{g}\right| \left(\sum_{n=1}^{N} \zeta(\mathbf{x}_n)l_{ng}\left(\mathbf{x}_{n}\right)+ \sum_{m=1}^{M} \varphi(\mathbf{y}_m)\hat{z}_{mg}\left(\mathbf{y}_{m}\right)\right) + \sum_{g=1}^{G} tr \left[\boldsymbol{\Sigma}^{-1}_{g}\left(\boldsymbol{W}_g^{X} + \boldsymbol{W}_g^{Y}\right) \right] \right]
\end{multline}
where $\boldsymbol{W}_g^{X}=\sum_{n=1}^{N} \zeta(\mathbf{x}_n)l_{ng}\left[ \left(\mathbf{x}_{n}-\boldsymbol{\mu}_{g}\right)\left(\mathbf{x}_{n}-\boldsymbol{\mu}_{g}\right)^{\prime}\right]$ and $\boldsymbol{W}_g^{Y}=\sum_{m=1}^{M} \varphi(\mathbf{y}_m)\hat{z}_{mg}\left[ \left(\mathbf{y}_{m}-\boldsymbol{\mu}_{g}\right)\left(\mathbf{y}_{m}-\boldsymbol{\mu}_{g}\right)^{\prime}\right]$. Finally, considering the eigenvalue decomposition $\boldsymbol{\Sigma}_g=\lambda_g\boldsymbol{D}_g\boldsymbol{A}_g\boldsymbol{D}^{'}_g$, \eqref{q_mstep_sigma} simplifies to:
\begin{align} \label{q_mstep_sigma_final}
\begin{split}
&-\frac{1}{2}\left[ \sum_{g=1}^{G} p \log \lambda_g \left(\sum_{n=1}^{N} \zeta(\mathbf{x}_n)l_{ng}\left(\mathbf{x}_{n}\right)+ \sum_{m=1}^{M} \varphi(\mathbf{y}_m)\hat{z}_{mg}\left(\mathbf{y}_{m}\right)\right) \right.+\\
&+ \left. \sum_{g=1}^{G} \frac{1}{\lambda_g} tr \left[\boldsymbol{D}_{g} \boldsymbol{A}^{-1} \boldsymbol{D}_{g}^{\prime} \left(\boldsymbol{W}_g^{X} + \boldsymbol{W}_g^{Y}\right) \right] \right]
\end{split}
\end{align}
The partial derivative of \eqref{q_mstep_sigma_final} with respect to $\left(\lambda_g,\boldsymbol{A}_g, \boldsymbol{D}_g\right)$ depends on the considered patterned structure: for a thorough derivation the reader is referred to \cite{Bensmail1996}. If \eqref{eigen_contr} is not satisfied, the constraints are enforced as detailed in Appendix C. Lastly, notice that in performing the concentration step the optimal observations of both training and test sets are retained, i.e. the ones with the highest contribution to the objective function.

The afore-described procedure falls within the structure of a general EM algorithm, for which the likelihood function does not decrease after an EM iteration, as shown in \cite{Dempster1977} and reported in page 78 of \cite{Gentle2006a}.
\begin{flushright}
$\square$
\end{flushright}

\section*{Appendix B}
This appendix details the structure of the Simulation Study in Section \ref{sim_setup_2}.
We consider a data generating process given by a mixture of $G=4$ components of multivariate t-distributions \citep{McLachlan2005, Peel2000}, according to the following parameters:
\[ \boldsymbol{\tau}=(0.2, 0.4, 0.1, 0.3)', \quad \nu=6, \]
\[ \boldsymbol{\mu}_1=(0, 0, 0, 0, 0, 0, 0,0,0,0,0)',\]
 \[ \boldsymbol{\mu}_2=(4, -4, 4, -4,4, -4,4, -4,4, -4)',\]
  \[ \boldsymbol{\mu}_3=(0,0,7,7,7,3,6,8,-4,-4)',\]
    \[ \boldsymbol{\mu}_4=(8, 0, 8, 0, 8, 0, 8,0,8,0,8)',\]
 \[ \boldsymbol{\Sigma}_1 = diag(1,1,1,1,1,1,1,1,1,1),\]
  \[ \boldsymbol{\Sigma}_2 = diag(2,2,2,2,2,2,2,2,2,2),\]
\[    \boldsymbol{\Sigma}_3 = \boldsymbol{\Sigma}_4=\begin{bmatrix}
     5.05 & 1.26 & -0.35 & -0.00 & -1.04 & -1.35 & 0.29 & 0.07 & 0.69 & 1.17 \\ 
  1.26 & 2.57 & 0.17 & 0.00 & 0.27 & 0.11 & 0.61 & 0.11 & 0.59 & 0.89 \\ 
  -0.35 & 0.17 & 6.74 & -0.00 & -0.26 & -0.31 & -0.01 & 0.00 & 0.08 & 0.14 \\ 
  -0.00 & 0.00 & -0.00 & 5.47 & -0.00 & -0.00 & 0.00 & 0.00 & 0.00 & 0.00 \\ 
  -1.04 & 0.27 & -0.26 & -0.00 & 6.80 & -0.76 & -0.12 & -0.01 & 0.09 & 0.21 \\ 
  -1.35 & 0.11 & -0.31 & -0.00 & -0.76 & 7.75 & -0.26 & -0.04 & -0.03 & 0.03 \\ 
  0.29 & 0.61 & -0.01 & 0.00 & -0.12 & -0.26 & 4.76 & 0.06 & 0.38 & 0.60 \\ 
  0.07 & 0.11 & 0.00 & 0.00 & -0.01 & -0.04 & 0.06 & 4.18 & 0.07 & 0.11 \\ 
  0.69 & 0.59 & 0.08 & 0.00 & 0.09 & -0.03 & 0.38 & 0.07 & 3.23 & 0.60 \\ 
  1.17 & 0.89 & 0.14 & 0.00 & 0.21 & 0.03 & 0.60 & 0.11 & 0.60 & 3.24 \\ 
    \end{bmatrix}.
\]
 \begin{figure}
 \centering
  \includegraphics[scale=.3]{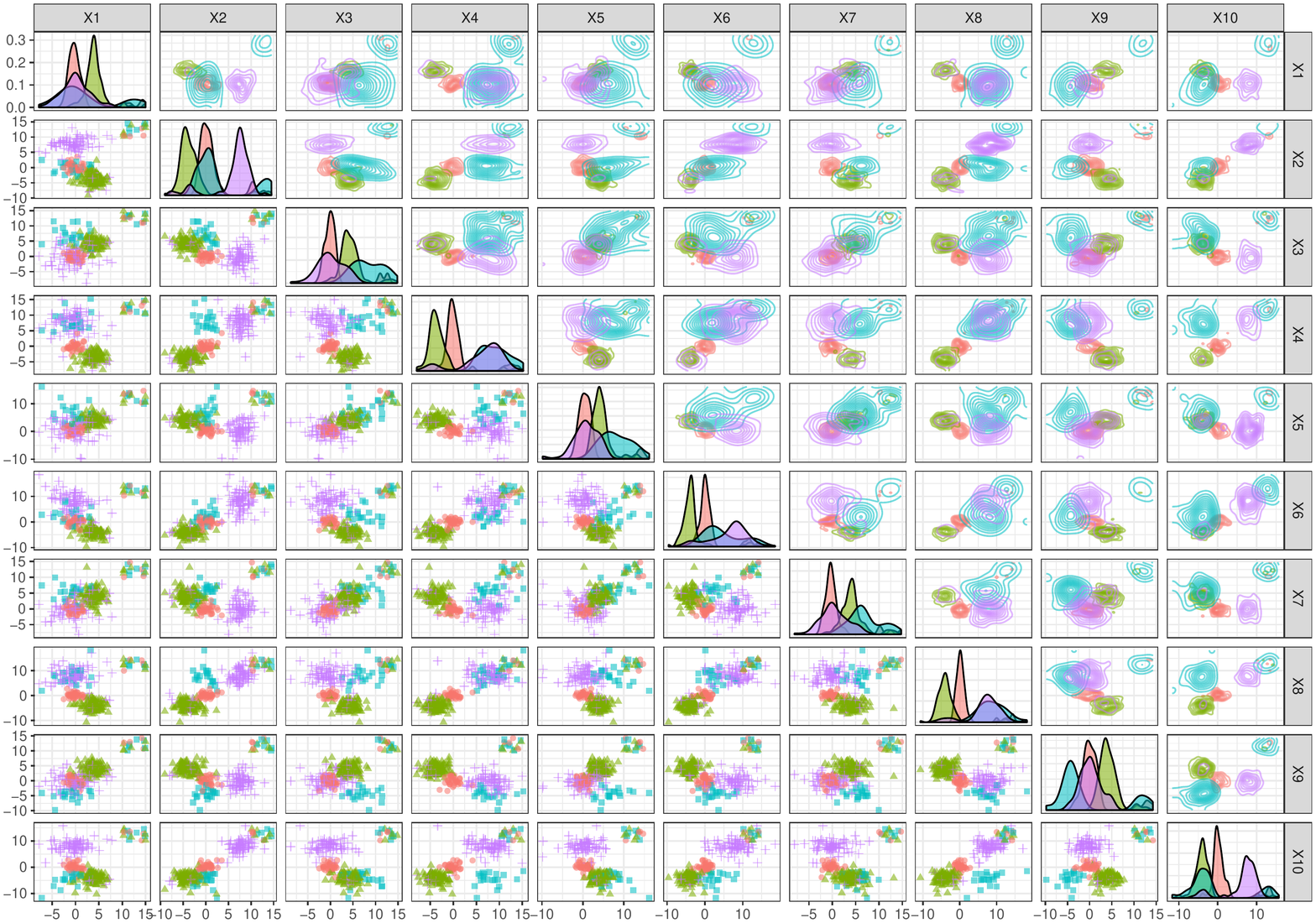}
\caption{Generalized pairs plot of the simulated data under the Simulation Setup described in \ref{sim_setup_2}. Both label noise and outliers are present in the data units.}
\label{fig:pair_plot}       % Give a unique label
\end{figure}   
A generalized pairs plot of contaminated labelled units under the afore-described Simulation Setup is reported in Figure \ref{fig:pair_plot}. %\textcolor{red}{Alternativamente, si pu\`{o} far vedere solo una porzione del pair plot, come in Figure \ref{fig:pair_plot_2}}
%\begin{figure}
% \centering
%  \includegraphics[scale=.3]{sim_study_2_pair_plot_small.pdf}
%\caption{Generalized pairs plot of the simulated data under the Simulation Setup described in \ref{sim_setup_2}. Both label noise and outliers are present in the data units.}
%\label{fig:pair_plot_2}       % Give a unique label
%\end{figure}
%The mixture model for the generic observation $\mathbf{w}_i \in \mathbb{R}^{10}$ has the following density:
%\begin{equation}
%f(\mathbf{w}_i,; \boldsymbol{\Theta}) =\sum_{g=1}^4  \tau_g f(\mathbf{w}_i; \boldsymbol{\mu}_g, \boldsymbol{\Sigma}_g, \nu)
%\end{equation}
%where each mixture component has a multivariate t-distribution:
%\begin{equation}
%f(\mathbf{w}_i; \boldsymbol{\mu}_g, \boldsymbol{\Sigma}_g, \nu)
%\end{equation}
\section*{Appendix C}
This final Section presents feasible and computationally efficient algorithms for enforcing the eigenvalue-ratio constraint according to the different patterned models in Table \ref{EDDA_model}.
%\subsection*{Notation}
At the $k-$th iteration of the M step, %consider the estimated scattering matrix for group $g$ given the current estimates of the unknown labels and mean vectors, that is:
%\begin{align} \label{W}
%\begin{split}
%\hat{\boldsymbol{W}}_g^{(k+1)}&=\Bigg[\sum_{n=1}^N \zeta(\mathbf{x}_n) \left(l_{ng} \mathbf{x}_n-\hat{\boldsymbol{\mu}}_g^{(k+1)}\right) \left(l_{ng}\mathbf{x}_n-\hat{\boldsymbol{\mu}}_g^{(k+1)}\right)^{'}+ \\
%  &+\sum_{m=1}^M \eta(\mathbf{y}_m) \left(\hat{z}_{mg}^{(k+1)}\mathbf{y}_m-\hat{\boldsymbol{\mu}}_g^{(k+1)}\right) \left(\hat{z}_{mg}^{(k+1)}\mathbf{y}_m-\hat{\boldsymbol{\mu}}_g^{(k+1)}\right)^{'}\Bigg]  
%\end{split}  
%\end{align}
%For ease of notation, we will drop the exponent indicating the current iteration of the EM algorithm, denoting the quantity in \eqref{W} with $\boldsymbol{W}_g$,
%and let $\hat{n}_g=\sum_{n=1}^N\zeta(\mathbf{x}_n)l_{ng}+\sum_{m=1}^M\eta(\mathbf{y}_m)\hat{z}_{mg}^{(k+1)}$ be the estimated group size. %$g=1,\ldots,G$ %with  $\boldsymbol{W}_g=\boldsymbol{L}_g \boldsymbol{\Omega}_g \boldsymbol{L}^{'}_g$ be its spectral decomposition.
the goal is to update the estimates %of the volume, shape and orientation
for the variance-covariance matrices $\hat{\boldsymbol{\Sigma}}_g^{(k+1)}=\hat{\lambda}_g^{(k+1)}\hat{\boldsymbol{D}}_g^{(k+1)}\hat{\boldsymbol{A}}_g^{(k+1)}\hat{\boldsymbol{D}}^{'(k+1)}_g$, $g=1,\ldots,G$ such that,
\begin{equation} \label{EIGEN_RESTR}
\frac{\max_{g=1\ldots G}\max_{l=1\ldots p}\hat{\lambda}_g^{(k+1)}\hat{a}_{lg}^{(k+1)}}
{\min_{g=1\ldots G}\min_{l=1\ldots p}\hat{\lambda}_g^{(k+1)}\hat{a}_{lg}^{(k+1)}} \leq c
\end{equation}
where $\hat{a}_{lg}^{(k+1)}$ indicates the diagonal entries of matrix $\hat{\boldsymbol{A}}_g^{(k+1)}$.
 %(exponent $(k+1)$ will be dropped from now on)
%according to the selected patterned model (see Table \ref{EDDA_model}) and eigenvalue-ratio constraint $c$.
Denote with $\hat{\Sigma}_g^{U}=\hat{\lambda}_g^{U} \hat{\boldsymbol{D}}_g^{U}\hat{\boldsymbol{A}}_g^{U}\hat{\boldsymbol{D}}_g^{'U}$ the estimates for the variance covariance matrices obtained following \cite{Bensmail1996} without enforcing the eigenvalues-ratio restriction in \eqref{EIGEN_RESTR}. Lastly, denote with $\hat{\boldsymbol{\Delta}}^U_g=\hat{\lambda}_g^{U}\hat{\boldsymbol{A}}_g^{U}$ the matrix of eigenvalues for $\hat{\boldsymbol{\Sigma}}_g^{U}$, with diagonal entries $\hat{d}_{lg}^U=\hat{\lambda}_g^{U}\hat{a}_{lg}^{U}$, $l=1,\ldots,p$.
%Consider a set of $G$ $p \times p$ positive semidefinite matrices $\boldsymbol{\Sigma}_g$, $g=1,\ldots,G$. Their spectral decomposition be:
%
%\begin{equation}
%\boldsymbol{\Sigma}_g= \boldsymbol{D}_g \boldsymbol{\Lambda}_g \boldsymbol{D}^{'}_g
%\end{equation}
%where $\boldsymbol{D}_g$ is a orthogonal matrix of eigenvectors and $\boldsymbol{\Lambda}_g=diag(d_{1g}, \ldots, d_{pg})$ where $\prod_{l=1}^pd_{lg}=|\boldsymbol{\Sigma}_g|$. Let us consider a further decomposition for $\boldsymbol{\Lambda}_g=\lambda_g \boldsymbol{A}_g$ where $\lambda_g$ is a scalar and $\boldsymbol{A}_g=diag(a_{1g}, \ldots, a_{pg})$ such that $\prod_{l=1}^p a_{lg}=1$. Notice that, by basic algebraic manipulation, $\lambda_g=|\boldsymbol{\Sigma}_g|^{1/p}$ and $d_{lg}=\lambda_g a_{lg}$.
%\vspace*{-.2cm}
\subsection*{Constrained maximization for VII, VVI and VVV models}
%\vspace*{-.3cm}
%Once having obtained optimal solutions for the selected model following \cite{Bensmail1996}, restricted eigenvalues satisfying \eqref{eigen_contr} are obtained by direct application of the \textit{optimal truncation operator} defined in \cite{Fritz2013a}. %We refer the reader to the aforementioned articles for details.
\begin{enumerate}
\item Compute $\boldsymbol{\Delta}_g$ applying the \textit{optimal truncation operator} defined in \cite{Fritz2013a} to $\left\{\hat{\boldsymbol{\Delta}}^U_1,\ldots,\hat{\boldsymbol{\Delta}}^U_G\right\}$, under condition \eqref{EIGEN_RESTR}%. The obtained matrices of eigenvalues $\boldsymbol{\Delta}_g$ satisfy \eqref{EIGEN_RESTR}
 \item Set  $\hat{\lambda}_g^{(k+1)}=|\boldsymbol{\Delta}_g|^{1/p}$, $\hat{\boldsymbol{A}}_g^{(k+1)}=\frac{1}{\hat{\lambda}_g^{(k+1)}}\boldsymbol{\Delta}_g$, $\hat{\boldsymbol{D}}_g^{(k+1)}=\hat{\boldsymbol{D}}_g^{U}$
\end{enumerate}
%\vspace*{-.5cm}
\subsection*{Constrained maximization for VVE model}
%\vspace*{-.3cm}
\begin{enumerate}
%\item Set $\boldsymbol{D}=\hat{\boldsymbol{D}}^U$
%\item Compute $\boldsymbol{\Delta}_g^*=\frac{1}{\hat{n}_g}diag\left(\boldsymbol{D}^{'}\hat{\boldsymbol{W}}_g^{(k+1)}\boldsymbol{D}\right)$
\item Compute $\boldsymbol{\Delta}_g$ applying the \textit{optimal truncation operator} defined in \cite{Fritz2013a} to $\left\{\hat{\boldsymbol{\Delta}}^U_1,\ldots,\hat{\boldsymbol{\Delta}}^U_G\right\}$, under condition \eqref{EIGEN_RESTR}
\item Given $\boldsymbol{\Delta}_g$, compute the common principal components $\boldsymbol{D}$ via, for example, a majorization-minimization (MM) algorithm \citep{Browne2014} %Use $\hat{\boldsymbol{D}}^{U}$ as initial starting value for the MM algorithm 
%\item Iterate $2-4$ until convergence
\item Set  $\hat{\lambda}_g^{(k+1)}=|\boldsymbol{\Delta}_g|^{1/p}$, $\hat{\boldsymbol{A}}_g^{(k+1)}=\frac{1}{\hat{\lambda}_g^{(k+1)}}\boldsymbol{\Delta}_g$, $\hat{\boldsymbol{D}}_g^{(k+1)}=\boldsymbol{D}$ 
\end{enumerate}
%\vspace*{-.9cm}
\subsection*{Constrained maximization for EVI, EVV models}
%\vspace*{-.3cm}
\begin{enumerate}
\item Compute $\boldsymbol{\Delta}_g$ applying the \textit{optimal truncation operator} defined in \cite{Fritz2013a} to $\left\{\hat{\boldsymbol{\Delta}}^U_1,\ldots,\hat{\boldsymbol{\Delta}}^U_G\right\}$, under condition \eqref{EIGEN_RESTR}
\item Compute $\boldsymbol{\Delta}^{\star}_g$ constraining $\boldsymbol{\Delta}_g$ such that $\boldsymbol{\Delta}^{\star}_g=\lambda^{\star}\boldsymbol{A}_g^{\star}$. That is, constraining $|\boldsymbol{\Delta}^{\star}_g|$ to be equal across groups \citep{Maronna1974,gallegos2002maximum}. Details are given in Section 3.2 of \cite{Fritz2012}
%\item Compute $\boldsymbol{\Delta}^{\star}_g$ applying the General Determinant Criterion \citep{Gallegos2002} to $\boldsymbol{\Delta}_g$, details are given in Section 3.2 of \cite{Fritz2012}. Notice that at this step $\boldsymbol{\Delta}^{\star}_g=\lambda_g^{\star}\boldsymbol{A}_g^{\star}$ with $\lambda_g^{\star}=\lambda^{\star}$, $ \forall g, g=1,\ldots, G$%, setting \texttt{restr.fact = 1}
\item Iterate $1-2$ until \eqref{EIGEN_RESTR} is satisfied
 \item Set  $\hat{\lambda}_g^{(k+1)}=\lambda^{\star}$, $\hat{\boldsymbol{A}}_g^{(k+1)}=\boldsymbol{A}_g^{\star}$, $\hat{\boldsymbol{D}}_g^{(k+1)}=\hat{\boldsymbol{D}}_g^{U}$
\end{enumerate}
%\vspace*{-.8cm}
\subsection*{Constrained maximization for EVE model}
%\vspace*{-.3cm}
\begin{enumerate}
\item Compute $\boldsymbol{\Delta}_g$ applying the \textit{optimal truncation operator} defined in \cite{Fritz2013a} to $\left\{\hat{\boldsymbol{\Delta}}^U_1,\ldots,\hat{\boldsymbol{\Delta}}^U_G\right\}$, under condition \eqref{EIGEN_RESTR}
\item Compute $\boldsymbol{\Delta}^{\star}_g$ constraining $\boldsymbol{\Delta}_g$ such that $\boldsymbol{\Delta}^{\star}_g=\lambda^{\star}\boldsymbol{A}_g^{\star}$. Details are given in Section 3.2 of \cite{Fritz2012}
\item Iterate $1-2$ until \eqref{EIGEN_RESTR} is satisfied
\item Given $\boldsymbol{A}^{\star}_g$, compute the common principal components $\boldsymbol{D}$ via, for example, a majorization-minimization (MM) algorithm \citep{Browne2014}
\item Set  $\hat{\lambda}_g^{(k+1)}=\lambda^{\star}$, $\hat{\boldsymbol{A}}_g^{(k+1)}=\boldsymbol{A}_g^{\star}$, $\hat{\boldsymbol{D}}_g^{(k+1)}=\boldsymbol{D}$
\end{enumerate}
%\vspace*{-.8cm}
\subsection*{Constrained maximization for VEI, VEV models}
%\vspace*{-.3cm}
\begin{enumerate}
\item Set $\boldsymbol{\Delta}_g=\hat{\boldsymbol{\Delta}}^U_g$
\item Set $\lambda_g^{\star}=\hat{\lambda}_g^{U}$, $g=1,\ldots,G$
\item Compute $\boldsymbol{\Delta}^{\star}_g$ applying the \textit{optimal truncation operator} defined in \cite{Fritz2013a} to $\left\{\boldsymbol{\Delta}_1,\ldots,\boldsymbol{\Delta}_G\right\}$, under condition \eqref{EIGEN_RESTR}
\item Compute $\boldsymbol{A}^{\star}=\left. \sum_{g=1}^G\frac{1}{\lambda_g^{\star}}\boldsymbol{\Delta}^{\star}_g \middle/ \left|\sum_{g=1}^G\frac{1}{\lambda_g^{\star}}\boldsymbol{\Delta}^{\star}_g \right|^{1/p} \right.$
\item Compute $\lambda_g^{\star}=\frac{1}{p}tr\left(\boldsymbol{\Delta}^{\star}_g {\boldsymbol{A}^{\star}}^{-1}\right)$
\item Set $\boldsymbol{\Delta}_g=\lambda_g^{\star}\boldsymbol{A}^{\star}$
\item Iterate $3-6$ until \eqref{EIGEN_RESTR} is satisfied
 \item Set  $\hat{\lambda}_g^{(k+1)}=\lambda^{\star}_g$, $\hat{\boldsymbol{A}}_g^{(k+1)}=\boldsymbol{A}^{\star}$, $\hat{\boldsymbol{D}}_g^{(k+1)}=\hat{\boldsymbol{D}}_g^{U}$
\end{enumerate}
%\vspace*{-.8cm}
\subsection*{Constrained maximization for VEE model}
%\vspace*{-.3cm}
\begin{enumerate}
\item Set $\boldsymbol{K}_g=\hat{\boldsymbol{\Sigma}}^U_g$
\item Set $\lambda_g^{\star}=\hat{\lambda}_g^{U}$, $g=1,\ldots,G$
\item Compute $\boldsymbol{K}_g^{\star}$ applying the \textit{optimal truncation operator} defined in \cite{Fritz2013a} to $\left\{ \boldsymbol{K}_1,\ldots,\boldsymbol{K}_G \right\}$, under condition \eqref{EIGEN_RESTR}
\item Compute $\boldsymbol{C}^{\star}=\left. \sum_{g=1}^G\frac{1}{\lambda_g^{\star}}\boldsymbol{K}^{\star}_g \middle/ \left|\sum_{g=1}^G\frac{1}{\lambda_g^{\star}}\boldsymbol{K}^{\star}_g \right|^{1/p} \right.$
\item Compute $\lambda_g^{\star}=\frac{1}{p}tr\left(\boldsymbol{K}^{\star}_g {\boldsymbol{C}^{\star}}^{-1}\right)$
\item Set $\boldsymbol{K}_g=\lambda_g^{\star}\boldsymbol{C}^{\star}$
\item Iterate $3-6$ until \eqref{EIGEN_RESTR} is satisfied
 \item Considering the spectral decomposition for $\boldsymbol{C}^{\star}=\boldsymbol{D}^{\star}\boldsymbol{A}^{\star}{\boldsymbol{D}^{\star}}^{'}$, set  $\hat{\lambda}_g^{(k+1)}=\lambda^{\star}_g$, $\hat{\boldsymbol{A}}_g^{(k+1)}=\boldsymbol{A}^{\star}$, $\hat{\boldsymbol{D}}_g^{(k+1)}=\boldsymbol{D}^{\star}$
\end{enumerate}

\end{document}